\documentclass[%
preprint,
amsmath,amssymb,
aps,
pre,
]{revtex4-2}

\usepackage[colorlinks,bookmarksopen,bookmarksnumbered,citecolor=red,urlcolor=red]{hyperref}

\usepackage{graphicx}
\usepackage{natbib}
\usepackage{color}
\usepackage{latexsym}
\usepackage{subfig}
\usepackage{natbib}
\usepackage{upgreek}
\usepackage{mathrsfs}
\usepackage{float}
\usepackage{caption}
\usepackage{soul}
\usepackage{textcomp}
\usepackage{stmaryrd}
\usepackage{amsmath}
\usepackage{lipsum}                     
\usepackage{xargs}                      
\usepackage[pdftex,dvipsnames]{xcolor}  
\usepackage[colorinlistoftodos,prependcaption,textsize=small]{todonotes}

\def\der#1#2{{\partial #1\over \partial #2}}

\newcommandx{\eyal}[2][1=]{\todo[linecolor=red,backgroundcolor=red!25,bordercolor=red,#1]{#2}}

\newcommandx{\anirban}[2][1=]{\todo[linecolor=blue,backgroundcolor=blue!25,bordercolor=blue,#1]{#2}}

\def\be{\begin{equation}}
\def\ee{\end{equation}}
\def\ee{{\rm e}}
\def\ii{{\rm i}}

\usepackage{cleveref}

\crefformat{section}{Sec. #2#1#3} 
\crefformat{subsection}{Sec. #2#1#3}
\crefformat{subsubsection}{Sec. #2#1#3}

\begin{document}

\preprint{APS/123-QED}

\title{On the normal form of  synchronization and resonance between vorticity waves in shear flow instability}

\author{Eyal Heifetz}
\affiliation{Department of Geophysics, Porter school of the Environment and Earth Sciences, Tel Aviv University,
Tel Aviv 69978, Israel.}

\author{Anirban Guha}%
 \email{anirbanguha.ubc@gmail.com}
\altaffiliation[Also at ]{Institute of Coastal Research, Helmholtz-Zentrum Geesthacht, Geesthacht 21502, Germany.}

\affiliation{Department of Mechanical Engineering, Indian Institute of Technology Kanpur, U.P. 208016, India}




\date{\today}

\begin{abstract}
A central mechanism of linearised two dimensional shear instability can be described in terms of a nonlinear, action-at-a-distance, phase-locking resonance between two vorticity waves which propagate  counter to their local mean flow as well as  counter to each other. Here we analyze the prototype of this interaction as an autonomous, nonlinear dynamical system. The wave interaction equations can be written in a generalized Hamiltonian action-angle form. The pseudo-energy serves as the Hamiltonian of the system, the action coordinates are the contribution of the vorticity waves to the wave-action, and the angles are the phases of the vorticity waves. The term ``generalized action-angle'' emphasizes that the action of each wave is generally time dependent, which allows instability. The synchronization mechanism between the wave phases depends on the cosine of their relative phase, rather than the sine as in the Kuramoto model. The unstable normal modes of the linearised dynamics correspond to the stable fixed points of the dynamical system and vice versa. Furthermore, the normal form of the wave interaction dynamics reveals a new type of inhomogeneous bifurcation -- annihilation of a pair of stable and unstable fixed points yields the emergence of two  neutral center fixed points of opposite circulation. 


\end{abstract}

\maketitle


\section{Introduction}
\label{intro}
\begin{figure}
\centering\includegraphics[width=0.8\linewidth]{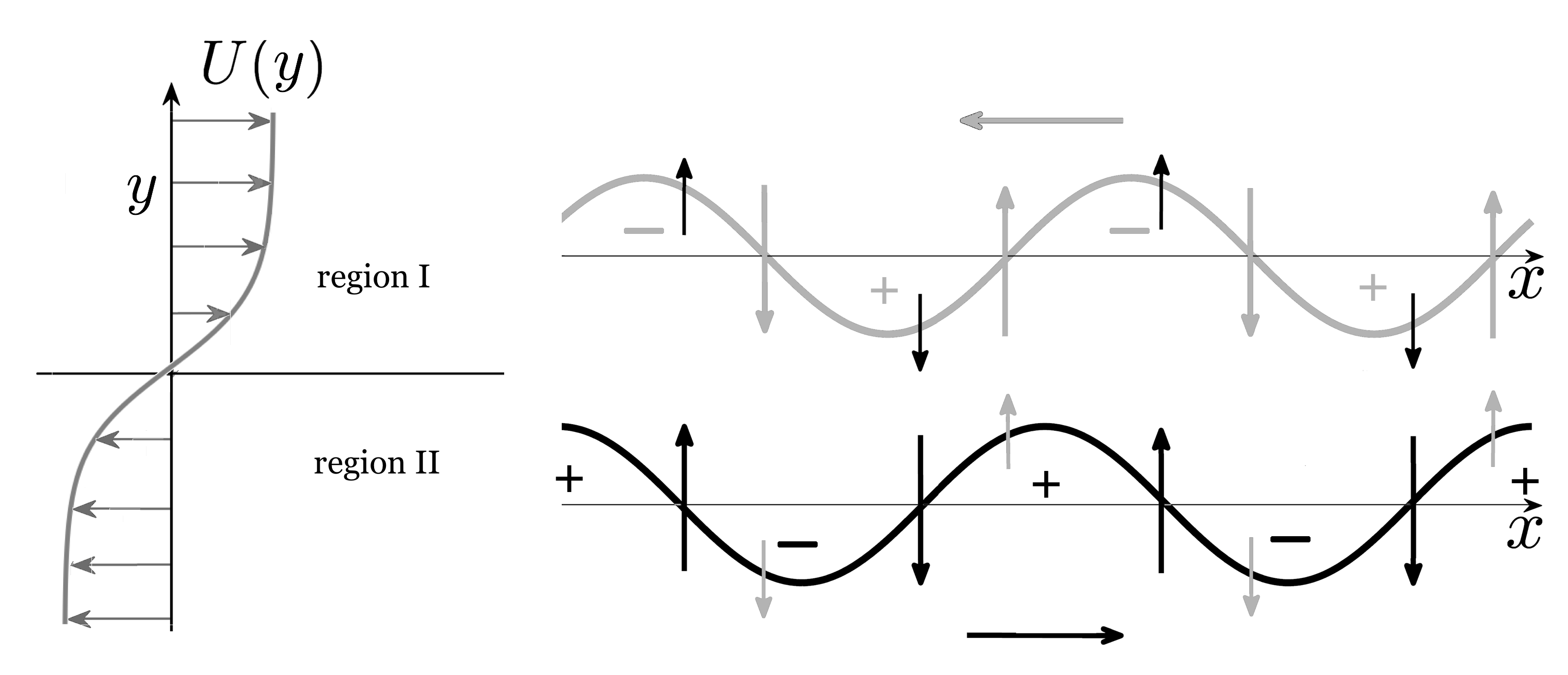}
\caption{Schematic of interacting vorticity waves in a shear flow. The  interfacial  displacement, the  associated  cross-stream  velocity  and the associated sign of  vorticity for  each  wave  are  shown  by  the  same  colour.  Interaction leads to  an  additional  cross-stream  velocity  (shown  by  a  different  colour).  The  horizontal  arrow associated  with  a  wave  indicates  the  intrinsic  wave  propagation  direction.  Both  waves are counter-propagating,  i.e.  moving  opposite  to  the  background  velocity  at  that  location.}
\label{fig:1}
\end{figure}

In many scenarios of two-dimensional shear flow instabilities, mainly in quasi 2D geophysical shear flows, the essence of the instability mechanism can be described in terms of a resonant interaction at a distance between two counter-propagating vorticity waves \citep{hosk1985,caul1994,heif1999,heif2005,carp2012,guha2014}. The schematic picture of the interaction can be drawn as follows. Consider two regions in the shear flow (indicated as regions  I and II   in Fig. \ref{fig:1}), where in region I the mean flow is positive, and in II it is negative.  Let us define the vorticity as positive when the circulation is counterclockwise and negative when it is clockwise. In the absence of a mean flow, a transversal vorticity wave will act to propagate to the left when the wave's  displacement and vorticity are in anti-phase (as in region I).

 This is because  the velocity induced by the vorticity field translates the displacement to its left. As the wave in  region I acts to propagate counter the mean flow, 
  it is denoted as a counter-propagating vorticity wave. 
Similarly, a counter-propagating wave in region II is obtained when the wave's vorticity and displacement are in phase. Examples of such vorticity waves are Rossby \citep{heif1999}, gravity \citep{harnik2008}, capillary \citep{biancofiore2015} and Alfven \citep{heifetzalfven2015} waves.

Although the wave's vorticity field may be localized, the velocity field attributed to this vorticity is non-local and decays away from the wave. Consequently, the two waves can interact at a distance by inducing on each other their individual cross-stream velocities. If the two waves' vorticity fields are in phase (Fig. \ref{fig:2}(a)) their cross-stream velocity will be in phase as well. Therefore, the induced velocity of one wave on the other will ``help'' the latter to translate its displacement faster and as a result, each wave will be propagating faster counter to its mean flow. In contrast, if the vorticity of the waves are in anti-phase (Fig. \ref{fig:2}(b)), the waves will hinder each other's counter-propagation rate. If the upper wave's vorticity lags the lower one by a quarter of a  wavelength (so that the waves are $\pi/2$ out of phase), the far field velocity induced by each wave will not affect the propagation rate but will amplify the waves' displacements. As each wave's displacement amplitude is tied to its vorticity, increase in the vorticity amplitude of one wave will lead to an amplification of the vorticity amplitude of the other wave. Therefore, this scenario describes a mutual instantaneous amplification at a distance (Fig. \ref{fig:2}(c)). In contrast, if it is the lower wave's vorticity which is lagging the upper one by a quarter of a wavelength, the waves will mutually decay each other's amplitudes (Fig. \ref{fig:2}(d)). Generally, any setup of phase difference between the two waves yields mutual interactions that affect both on the waves' amplitudes and the waves' propagation rates (Fig. \ref{fig:2}(e)).

The wave interaction picture described above can be translated into a generic  set of equations. On writing the vorticity waves at the two regions in terms of their vorticity amplitudes and phases we get:
\begin{figure}
\centering\includegraphics[width=0.9\textwidth]{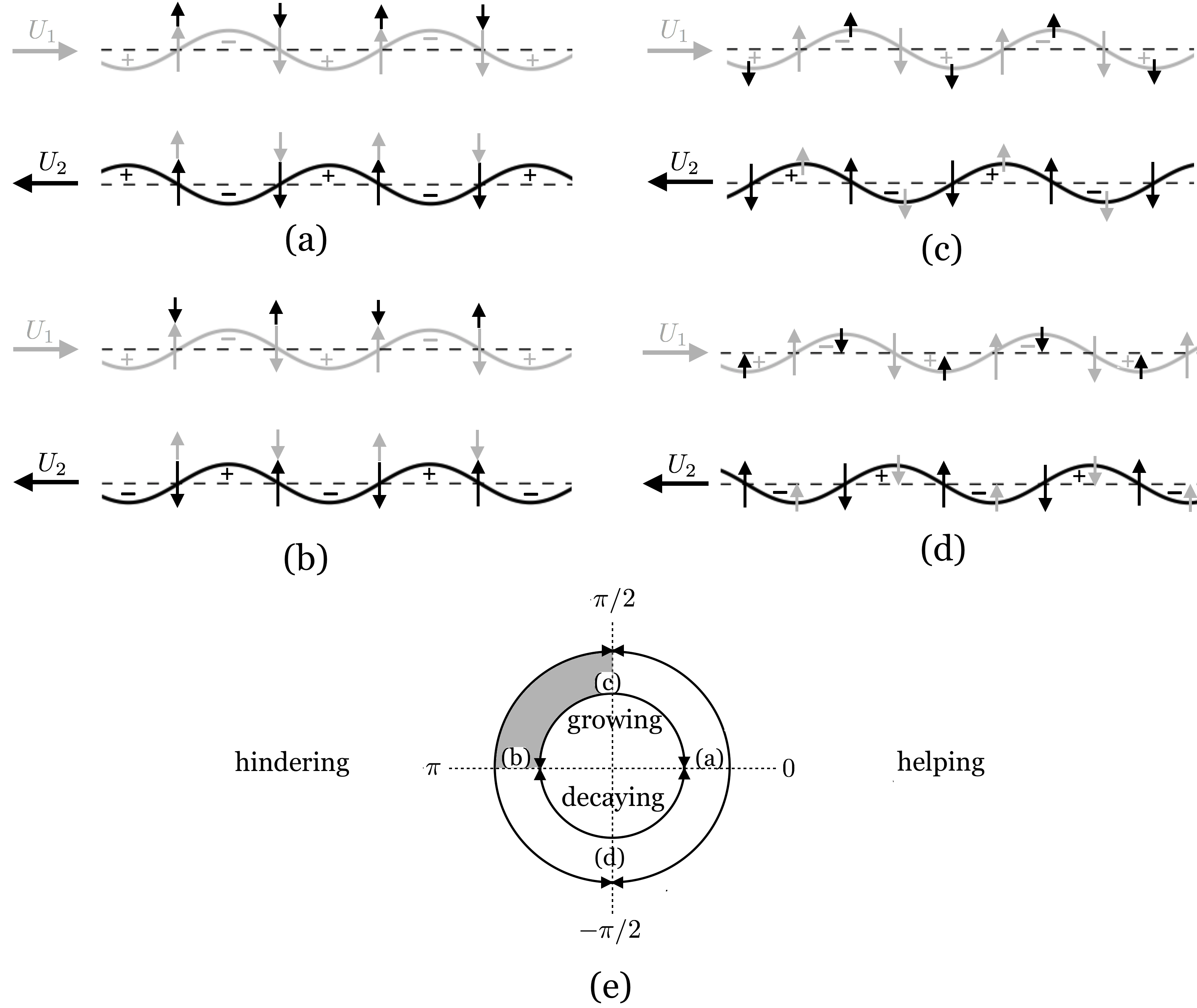}
\caption{Schematic description of the linear interactions between counter-propagating vorticity waves. The waves depict interfacial displacement, while the horizontal and vertical arrows  respectively denote streamwise (background) and cross-stream velocities. (a) Fully helping, (b) fully hindering, (c) fully growing and (d) fully decaying configurations. (e) Depending on the phase difference $\epsilon$ between the vorticity perturbations at the upper and lower interface, different kind of linear interactions can be expected, as shown by the `concentric circles'. The locations where the configurations (a)-(d) occur have been marked. The configuration given in Fig. \ref{fig:1} lies in the second quadrant (shaded in grey), which is the `growing-hindering configuration'.} 
\label{fig:2}
\end{figure}
$q_{1,2} = Q_{1,2}\ee^{\ii\epsilon_{1,2}}$,  the vorticity amplitude and phase equations become: 
\begin{subequations}\label{eq:1.1}
\begin{align}
{\dot Q}_{1} = \sigma_{1}Q_{2}\sin{\epsilon}, \qquad {\dot Q}_{2} = \sigma_{2}Q_{1}\sin{\epsilon},  \\
{\dot \epsilon}_{1}  = - {\hat \omega}_{1} + \sigma_{1}{Q_{2}\over Q_{1}} \cos{\epsilon},\qquad {\dot \epsilon}_{2}  = - {\hat \omega}_{2} - \sigma_{2}{Q_{1}\over Q_{2}} \cos{\epsilon},
\end{align}
\end{subequations}
where $\epsilon \equiv \epsilon_1 - \epsilon_2$ is the waves' relative phase, ${\hat \omega}_{1,2}$ are the waves' frequencies in the absence of interaction, $\sigma_{1,2}$ are the interaction at a distance coefficients, which depend on the details of the problem but usually increases monotonically with wavelength. Thus, Eq. (\ref{eq:1.1})a indicates that indeed $\epsilon = \pi/2$ (Fig. \ref{fig:2}(c))  is  the optimal phase for mutual instantaneous amplification.   
 The frequencies ${\hat \omega}_{1,2}$ include both the effects of advection by the mean flow, $U_{1,2}$ (which provides the Doppler shift) and the counter-propagation rate (which is the intrinsic frequency).  The frequency of each wave can be either positive or negative. If, for instance,  ${\hat \omega}_1 > 0$, it implies that in the absence of interaction, the wave's phase speed is directed to the right. Therefore, the counter-propagation rate of wave `1' is not strong enough to overcome the local mean flow. Generally,  wave counter-propagation acts to increase $\epsilon_1$ and decrease $\epsilon_2$. Hence if, for instance, the waves are in phase  ($\cos{\epsilon}=1$, see Fig. \ref{fig:2}(a))\footnote{Note that in Fig. \ref{fig:2}, the displacement fields have been plotted. In Fig. \ref{fig:2}(a) the displacement fields are anti-phased however the vorticity fields are in phase.}, the waves help each other to counter-propagate in agreement with the $+$ and $-$ signs associated with the last terms in Eq.  (\ref{eq:1.1}(b)). It is worth noting that although the system Eq.  \eqref{eq:1.1} is composed of inhomogeneous \emph{nonlinear} ordinary differential equations, it describes the \emph{linearized dynamics} of small perturbations in shear flows. This is, to some extent,  analogous to the nonlinear description of the small amplitude dynamics of two coupled pendulum. In this paper we explore the dynamical behavior of the system given by Eq.  \eqref{eq:1.1}.   

The organization of the paper is as follows. In \cref{Section 2}, we derive its general properties and then, in  \cref{Section 3}, we analyze the dynamics in details with relations to the physical interaction mechanism described in the introduction.
In  \cref{Section 4} we implement these results to a concrete example of a piecewise linear shear flow and  then in \cref{Section 5}, we generalize the two-wave interactions of Eq.  \eqref{eq:1.1} to the case of multiple-wave interactions. We close by concluding our results in \cref{Section 6}.  

\section{General dynamical properties} 
\label{Section 2}

\subsection{Generalized canonical action-angle formulation}
\label{Section 2a}
It is straightforward to verify that the system Eq.  \eqref{eq:1.1} conserves the following two constant of motion:

\begin{subequations}\label{eq:2.1}
\begin{align}
{\cal H} & = -{\hat \omega}_1 {Q_1^2\over 2\sigma_1} + {\hat \omega}_2{Q_2^2\over 2\sigma_2}+ Q_{1}Q_{2} \cos{\epsilon}  \nonumber\\
& = -({\hat \omega}_1 {\cal A}_1+ {\hat \omega}_1 {\cal A}_2 ) + 2\ii\sigma\sqrt{{\cal A}_1{\cal A}_2}\cos{\epsilon},\\
{\cal A} & = {Q_1^2\over 2\sigma_1} -{Q_2^2\over 2\sigma_2} = {\cal A}_1 + {\cal A}_2 ,
\end{align}
\end{subequations}
where $\sigma \equiv \sqrt{\sigma_1 \sigma_2}$ is the geometric mean of the interaction coefficients.
The first conserved quantity is denoted as the pseudo-energy of the system and the second is its action, which is also proportional to the psuedo-momentum  \cite{heifetz2004counter,heifetz2018generalized}. The term `generalized' is used here because, as opposed to the classical action-angle formulation, the actions associated with each wave (each degree of freedom): ${\cal A}_{1} = [{Q^2 /(2\sigma)}]_{1}$ and ${\cal A}_{2} = -[{Q^2 /(2\sigma)}]_{2}$  are not conserved individually.  It is also straightforward to verify that $${\cal H} = {\cal A}_1 {\dot \epsilon}_1 + {\cal A}_2 {\dot \epsilon}_2.$$ Therefore,  Eqs. (\ref{eq:1.1}a)--(\ref{eq:1.1}b) can be rewritten in a canonical generalized action-angle form, in which ${\cal H}$ serves as the Hamiltonian:
\begin{subequations}\label{eq:2.2}
\begin{align}
{\dot {\cal A}}_{1} = 2 \ii\sigma\sqrt{{\cal A}_1{\cal A}_2}\sin{\epsilon}  = -\der{{\cal H}}{\epsilon_{1}},\nonumber \\
{\dot {\cal A}}_{2} = -2 \ii\sigma\sqrt{{\cal A}_1{\cal A}_2}\sin{\epsilon}  = -\der{{\cal H}}{\epsilon_{2}},\\
{\dot \epsilon}_{1} = - {\hat \omega}_{1} + \ii\sigma\sqrt{{\cal A}_{2}\over {\cal A}_{1}}\cos{\epsilon} = \der{{\cal H}}{{\cal A}_{1}}, \nonumber\\ 
{\dot \epsilon}_{2} = - {\hat \omega}_{2} + \ii\sigma\sqrt{{\cal A}_{1}\over {\cal A}_{2}}\cos{\epsilon} = \der{{\cal H}}{{\cal A}_{2}}.
\end{align}
\end{subequations}
To avoid confusion, please note that Eqs. (\ref{eq:2.1})--(\ref{eq:2.2}) comprise only of real terms as ${\cal A}_{2}$ is negative definite. 

\subsection{The complex normal form}

The nonlinear, real, inhomogeneous  set of Eqs. (\ref{eq:1.1}a)--(\ref{eq:1.1}b) results from the set of two linear, complex, homogeneous equations  (see Eqs. (9a)--(9c) of Ref. \cite{heif2005} and Eqs. (4.1)--(4.2) of Ref. \cite{guha2014}):
\begin{subequations}\label{eq:2.3} 
\begin{align}
{\dot q}_{1} = -\ii{\hat \omega}_{1}q_1 +\ii\sigma_{1}q_{2}\, ,\\
{\dot q}_{2} = -\ii{\hat \omega}_{2}q_2 -\ii\sigma_{2}q_{1}\, .
\end{align}
\end{subequations}
Define the complex variable:
\begin{equation}\label{eq:2.4}
{\cal Z}\equiv \sqrt{\sigma_2\over \sigma_1}{q_1\over q_2}\equiv \chi \ee^{\ii\epsilon}\, ,
\end{equation}
where $\chi = \sqrt{\sigma_2/\sigma_1}({Q_1 /Q_2})$ is a scaled ratio of the wave amplitudes. Hence Eqs. (\ref{eq:2.3}a)--(\ref{eq:2.3}b) can be then written in the compact complex form:
\begin{equation}\label{eq:2.5}
{\dot {\cal Z}} = \ii\sigma\left [{\cal Z}\left ({\cal Z} - {{\hat \omega}\over \sigma} \right ) +1 \right ]\, , 
\end{equation}
where ${\hat \omega} \equiv {\hat \omega}_1 - {\hat \omega}_2$. Defining the control parameter $\mu \equiv {\hat \omega}/\sigma$,  and using a scaled time $\tau \equiv \sigma t$,  Eq.  \eqref{eq:2.5} can be expressed as the following normal form:
\begin{equation}\label{eq:2.6}
{d{\cal Z} \over d\tau} =  \ii\left [{\cal Z}\left ({\cal Z} - \mu\right ) +1 \right ]\, . \end{equation}
The normal form is complex as it describes the evolution of both the waves' amplitudes and phases. It is inhomogenous since the mean flow acts as an external forcing. Furthermore, the essence of the system dynamics is an interaction between waves with different intrinsic frequencies. Thus, the dynamics is controlled by a single parameter, which is the ratio between the frequency difference of the  two waves and the mean interaction coefficient.

\subsection{Relations between fixed points and normal modes}

The complex fixed points of Eq.  \eqref{eq:2.6} are obtained when  the waves' (scaled) amplitude ratio $\chi^*$ and their relative phase $\epsilon^*$ remain fixed (where the asterisk denotes the values at the fixed points):
\begin{equation}\label{eq:2.7}
{d{\cal Z}\over d\tau} = 
 \ee^{\ii\epsilon}\left [ {d\chi \over d\tau} + 
 \ii\chi {d \epsilon\over d\tau}\right ] = 0\, .
\end{equation}
We also note that these fixed points are the normal modes of the equivalent linear system Eq.  \eqref{eq:2.3}. To show this we write the latter in a matrix form:
\begin{align}\label{eq:2.8}
{\dot {\bf q}} = {\bf  A q}\, & ; \hspace{0.25cm}
{\bf q} = \left[\begin{array}{c}
q_1\\
q_2
\end{array}\right]\, ;\nonumber \\ 
{\bf A} = -\ii  \left[\begin{array}{cc}
{\hat \omega}_1  & -\sigma_1\\
\sigma_2 & {\hat \omega}_2
\end{array}\right]\, & \implies
 {\bf q} = \sum_{j=1}^2 a_j {\bf p}_j \ee^{\lambda_j t}\, ,
\end{align}
where $a_j$s are constants,  ${\bf p}_j$ and $\lambda_j$ are respectively the complex eigenvectors and eigenvalues of ${\bf A}$. Then if we denote:  
\begin{equation}\label{eq:2.9}
{\bf p}_j = \left[\begin{array}{c}
P_1 \ee^{\ii\phi_1}\\
P_2 \ee^{\ii\phi_2}
\end{array}\right]_j\, ; \hspace{0.25cm} 
\phi \equiv \phi_1 - \phi_2\, ; \hspace{0.25cm}
\lambda = \lambda_r +  \ii \lambda_i\, ,
\end{equation}
then the $j^{th}$ normal mode solution can be written as:
\begin{equation}\label{eq:2.10}
{\bf q} = 
\left[\begin{array}{c}
Q_1 \ee^{\ii\epsilon_1}\\
Q_2 \ee^{\ii\epsilon_2}
\end{array}\right]=
\left[\begin{array}{c}
(P_1\ee^{\lambda_r t}) \ee^{\ii(\phi_1 + \lambda_i t})\\
(P_2\ee^{\lambda_r t}) \ee^{\ii(\phi_2 + \lambda_i t})
\end{array}\right]_j\,.
\end{equation}
\begin{figure}
\centering\includegraphics[width=0.6\linewidth]{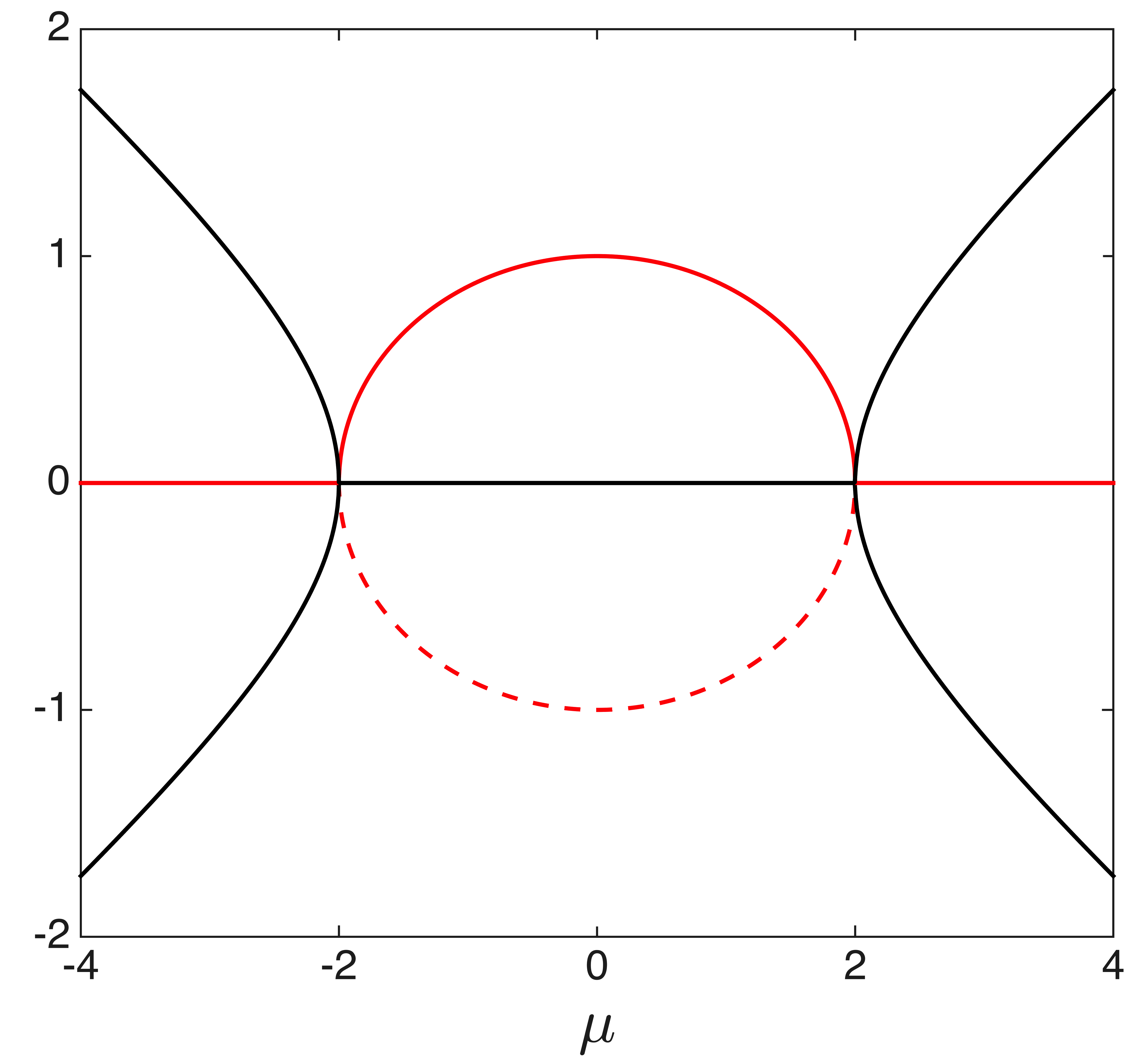}
\caption{Bifurcation diagram. Black line denotes $\lambda_i$ while red line denotes $\lambda_r$. The unstable normal modes ($\lambda_r > 0$) compose the branch of stable fixed points (marked by the solid red line) of the dynamical system Eq.  \eqref{eq:2.6}, whereas the stable modes ($\lambda_r < 0$) compose the unstable branch (marked by the dashed red line).}
\label{fig:lambda_mu}
\end{figure}

\noindent Therefore, $\chi^*=\sqrt{\sigma_2/\sigma_1}({Q_1 /Q_2})_j^* = \sqrt{\sigma_2/\sigma_1}({P_1 /P_2})_j = \mathrm{const_1}$ and $\epsilon_j^* = ({\epsilon_1 - \epsilon_2})_j = \phi_j = \mathrm{const_2}$. 
To simplify the analysis we hereafter refer to the motion in the frame of reference of the mean frequency (rather, phase speed) in the absence of interaction, i.e.\
${\overline \omega} = ({\hat \omega}_1 + {\hat \omega}_2)/2$. There, the eigenvalues of ${\bf A}$ (normalized by $\sigma$) satisfy:
\begin{equation}\label{eq:2.11}
\lambda^{NM}_{1,2} = \pm \sqrt{1- \left ({\mu \over 2}\right )^2}\,. 
\end{equation}
Note that the condition of constant wave amplitude ratio applies for normal modes with either positive, negative or zero growth rate ($\lambda^{NM}_r$), and the phase-locking condition implies that the waves are moving in concert with the same frequency ($-\lambda^{NM}_i$), which as well can be either positive, negative or zero. The dependence of eigenvalues on the bifurcation parameter, $\mu$ is shown in Fig.\, \ref{fig:lambda_mu}. Clearly, the nature of the eigenvalues change at $|\mu/2|=1$; the normal form Eq.  \eqref{eq:2.6} demonstrates a new kind of bifurcation where a pair of eigenvalues transform from pure real to pure imaginary. In the following section, we carry out this analysis in more detail and augment it by drawing the phase portrait of the system. The final aim is to link the dynamics on the phase plane with the mechanistic understanding of  wave interaction presented in the introduction.


\section{Dynamical system analysis}
\label{Section 3}
\subsection{Dynamics on a compact non-Hamiltonian degenerated phase plane}
\label{subsec:3compact}
\begin{figure}
\centering
\includegraphics[width=0.9\linewidth]{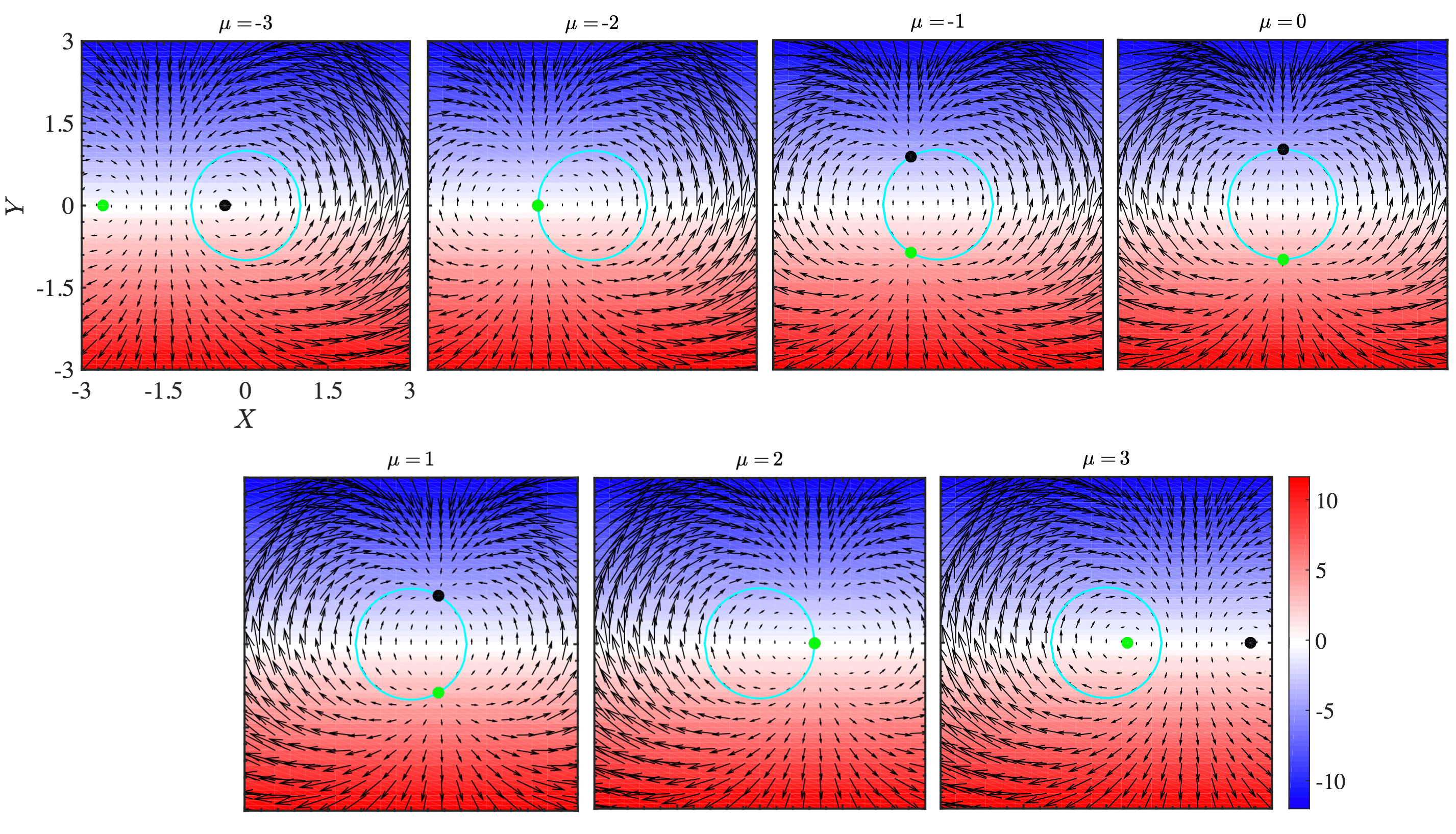}
\caption{Phase portrait with colors indicating the divergence field, $\mathcal{D}$. The green and black dots denote the fixed points. A unit circle, centered at the origin, is plotted in cyan.}
\label{fig:div_Phase_port}
\end{figure}

\begin{figure}
\centering\includegraphics[width=0.9\linewidth]{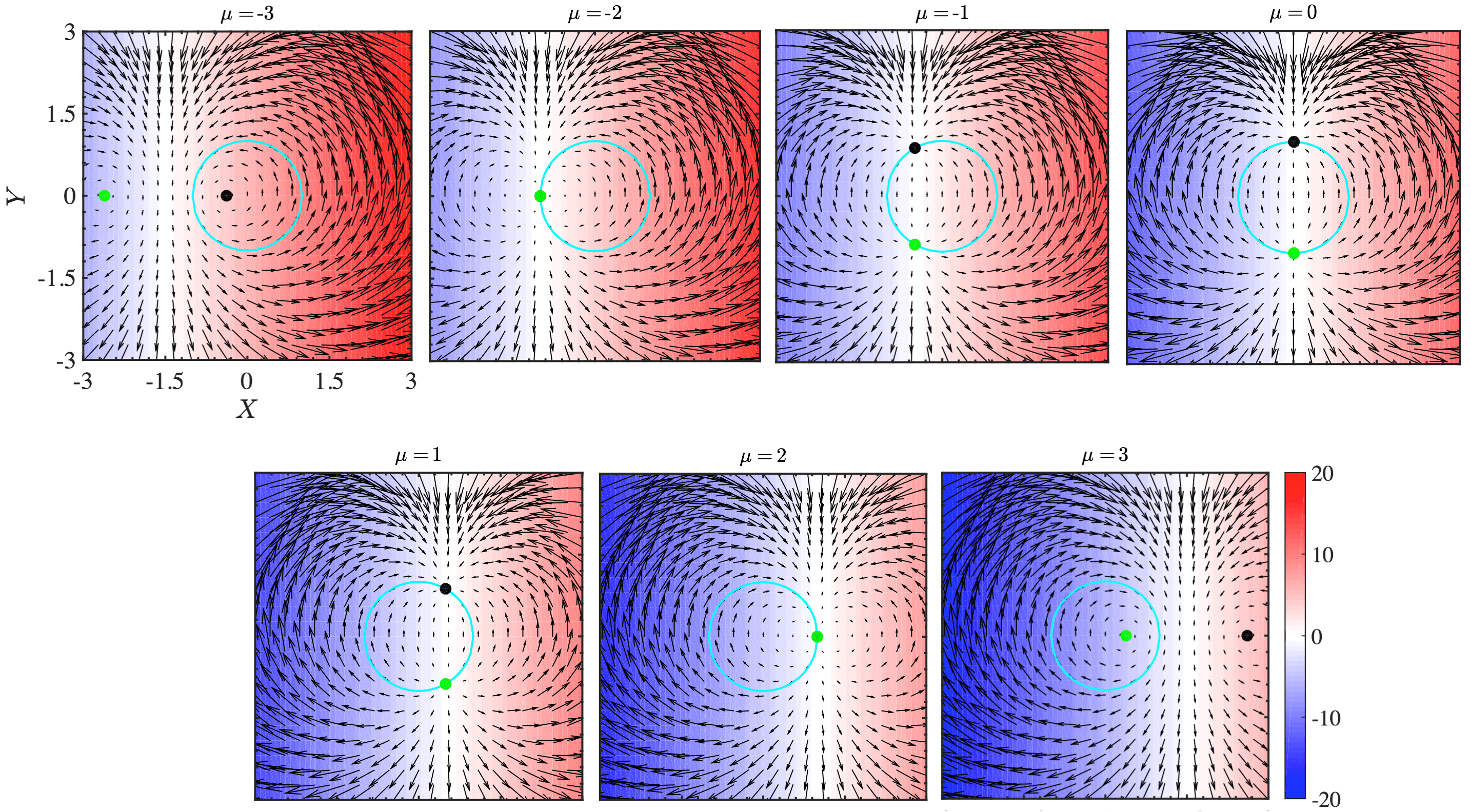}
\caption{Phase portrait with colors indicating the curl field, $\mathcal{C}$.}
\label{fig::curl_Phase_port}
\end{figure}

We can express Eq.  \eqref{eq:2.6} in terms of $(\chi, \epsilon)$ to obtain the following autonomous, nonlinear dynamical system:
\begin{subequations}\label{eq:3.1}
 \begin{align}
{d\chi\over d\tau} & = \left(1- \chi^2 \right) \sin \epsilon\, , \label{eq:polar_dynsyn1}
\\
{d\epsilon\over d\tau}  & = \left( \chi+{\chi}^{-1}\right)  \cos\epsilon -\mu\,.
\end{align}
\end{subequations}
We note that the (scaled) waves' amplitude ratio and phase difference are respectively within the ranges of $\chi\in(0,\,\infty)$ and $\epsilon\in[-\pi,\,\pi]$.
Equations (\ref{eq:3.1}a)--(\ref{eq:3.1}b) are in polar coordinates, with $\chi$ being the radius and $\epsilon$ being the azimuthal angle. Equivalently it can be expressed in a Cartesian form:
\begin{subequations}\label{eq:3.2}
 \begin{align}
\mathcal{U} \equiv {d X\over d\tau} & = Y(\mu -2 X) , \\
\mathcal{V} \equiv {d Y\over d\tau}  & = -\mu X +X^2-Y^2+1,
\end{align}
\end{subequations}
(where  $X=\chi \cos\epsilon$ and $Y=\chi \sin\epsilon$) from which we compute the divergence and curl of the phase plane flow:
\begin{subequations}\label{eq:3.3}
 \begin{align}
\mathcal{D}&\equiv \der{\mathcal{U}}{X} + \der{\mathcal{V}}{Y} = -4Y,\\
\mathcal{C}& \equiv \der{\mathcal{V}}{X} - \der{\mathcal{U}}{Y} =  -2(\mu-2X).
\end{align}
\end{subequations}
The phase portrait corresponding along with the divergence field, and the same along with the curl field are respectively shown in Fig.\ \ref{fig:div_Phase_port} and Fig.\      \ref{fig::curl_Phase_port}.
The fixed points in polar coordinates are:
\begin{subequations}\label{eq:3.4}
 \begin{align}
(\chi,\epsilon)^* & = \left(1,\,\pm \cos^{-1}\left({\mu \over 2}\right)\right)\,\,\mathrm{when}\,\,\left|{\mu \over 2}\right| \leq 1,\\
(\chi,\epsilon)^*  & = \left({\mu \over 2} \pm \sqrt{\left ({\mu \over 2}\right )^2 -1},\,0\right)\,\,\mathrm{when}\,\,{\mu \over 2} \geq 1\\
(\chi,\epsilon)^*  & = \left(-{\mu \over 2} \pm \sqrt{\left ({\mu \over 2}\right )^2 -1},\,\pi\right)\,\,\mathrm{when}\,\,{\mu \over 2} \leq -1,
\end{align}
\end{subequations}
or equivalently in Cartesian coordinates:
\begin{subequations}\label{eq:3.5}
 \begin{align}
(X,Y)^* & = \left({\mu \over 2},\,  \pm \sqrt{1-\left ({\mu \over 2}\right )^2 }\right)\,\,\mathrm{when}\,\,\left|{\mu \over 2}\right| \leq 1,\\
(X,Y)^*  & = \left({\mu \over 2} \pm \sqrt{\left ({\mu \over 2}\right )^2 -1},\,0\right)\,\,\mathrm{when}\,\,\left|{\mu \over 2}\right| \geq 1.
\end{align}
\end{subequations}
The stability of the fixed points is obtained from the eigenvalues, $\lambda_{1,2}^J$, of the Jacobian matrix $J$, evaluated at the fixed points: 
\begin{equation}\label{eq:J3.6}
\frac{d}{d\tau}\left[\begin{array}{c}
\delta X\\
\delta Y
\end{array}\right]={1\over 2}\left[\begin{array}{cc}
\mathcal{D} & -\mathcal{C}\\
\mathcal{C} & \mathcal{D}
\end{array}\right]^*\left[\begin{array}{c}
\delta X\\
\delta Y
\end{array}\right],
\end{equation}
yielding $\lambda_{1,2}^J = (\mathcal{D}^* \pm i\mathcal{C}^*)/2$. Transforming $(\delta X, \delta Y) =\delta R(\cos{\theta}, \sin{\theta})$ to polar coordinates whose origin is located at the fixed points, 
we obtain at the fixed points' vicinity:
\begin{equation}\label{eq:3.7}
\delta R = \delta R_0e^{\mathcal{D}^*\tau/2}; \hspace{0.5cm} \delta \theta = \delta \theta_0 + \mathcal{C}^*\tau/2.
\end{equation}

\subsubsection{The case of $\, |{\mu \over 2}| < 1$}

The control parameter $\mu$ represents the ratio between the difference between the waves' frequencies (in the absence of interaction) and the mean interaction coefficient. The former acts to shear the waves apart, whereas the latter acts to keep them together. 

When $\, |{\mu/ 2}| < 1$, this ratio is not very large and the obtained fixed points are located on the unit circle ($\chi = 1$),  see Eq.  (\ref{eq:3.4}a).
We first note that on the unit circle, the total wave action ${\cal A}$ given in Eq.  (\ref{eq:2.1}b) vanishes. Furthermore, at the fixed points, when in addition $\cos{\epsilon} = {\mu/ 2}$, the pseudo-energy, given in Eq.  (\ref{eq:2.1}a), vanishes as well. This allows normal mode exponential growth ($\lambda_r^{NM} > 0$) or decay ($\lambda_r^{NM} < 0$) since then these two constants of motion remain zero despite the temporal change in the waves' amplitudes \footnote{In order to see that substitute Eq.  \eqref{eq:2.10} in Eqs. (\ref{eq:2.1}a)--(\ref{eq:2.1}b) to obtain :
${\cal H}  = \left[-{\hat \omega}_1 {P_1^2\over 2\sigma_1} + {\hat \omega}_2{P_2^2\over 2\sigma_2}+{\mu \over 2} P_{1}P_{2} \right]
e^{2\lambda_r t}\,$;

${\cal A}  = \left({P_1^2\over 2\sigma_1} -{P_2^2\over 2\sigma_2}\right)e^{2\lambda_r t}$. Therefore both $({\cal H}, {\cal A})$  must vanish to remain constant when $\lambda_r \neq 0$. }. On the unit circle, the Equation-set (\ref{eq:3.1}a)--(\ref{eq:3.1}b)  reduces to:
\begin{equation}\label{eq:3.6}
{d\epsilon\over d{\tilde \tau}}  = \cos\epsilon -{\mu\over 2}\, ,  
\end{equation}
where ${\tilde \tau} = 2\tau$. Comparing with the synchronization model of  Kuramoto  \cite{PIK01},  here in the RHS, we have cosine of the phase difference rather than its sine. This is because the nature of the wave interaction is fundamentally different from the one described in the Kuramoto model. As discussed in \cref{intro} and specifically in Fig.\, \ref{fig:2}, each wave does not try to adjust its frequency to the other to obtain synchronization. In contrast, when they are in phase, they act to increase their phase difference, see Fig. \ref{fig:2}(a), and when they are in quadrature ($\pi/2$ out of phase), the phase difference is not affected at all by the wave interaction, see Fig.\, \ref{fig:2}(c).


Furthermore, since $\epsilon^*=\pm \cos^{-1}(\mu/2)$, it implies that the fixed points are symmetric with respect to $\epsilon=  0$ (this translates to a reflection symmetry about the $X$-axis in Figs. \ref{fig:div_Phase_port} and \ref{fig::curl_Phase_port}). Consulting with Fig.\ \ref{fig:2}, growth is obtained when $0<\epsilon <\pi$ (upper half-plane of Figs. \ref{fig:div_Phase_port} and \ref{fig::curl_Phase_port}) and decay when $-\pi<\epsilon < 0$ (lower half plane). When $\mu = 0$, the waves propagate in concert in the absence of interaction. Hence the only way to keep them locked in the presence of interaction is to prevent the interaction to affect the waves' phase speeds. Thus, the waves' phase difference is either $\pi/2$ (for amplitude growth, Fig.\ \ref{fig:2}(c)) or $-\pi/2$ (for amplitude decay, Fig.\ \ref{fig:2}(d)). For positive $\mu$, the waves should help each other to counter-propagate in order to remain phase-locked ($-\pi/2<\epsilon < \pi/2$).  In contrast, when $\mu$ is negative, the  waves should hinder their counter-propagation rate ($\pi/2<\epsilon < 3\pi/2$). 

The unstable (stable) normal modes are obtained when the amplitude of the two waves grow (decay) with the same exponential growth (decay) rate  $\lambda_r^{NM}$. Indeed, as indicated from Eq.  (\ref{eq:1.1}a) and Eq.  \eqref{eq:2.11}, for $\chi^*=1$:
\begin{equation}
\lambda_r^{NM} = {1\over \sigma}{\dot{Q}_1 \over Q_1} = {1\over \sigma}{\dot{Q}_2 \over Q_2} = \sin{\epsilon^*},
\end{equation}
which is positive (negative) for the upper (lower) part of the inner circle in Fig.\, \ref{fig:2}(e). Furthermore, in the frame of reference moving with the mean frequency ${\overline\omega}$, Eq.  \eqref{eq:2.11} also yields:  
\begin{equation}
\lambda_i^{NM} = {(\dot{\epsilon}_1)^* \over \sigma} = {(\dot{\epsilon}_2)^*  \over \sigma}= 0,
\end{equation}
hence for $|{\mu / 2}| < 1$, we obtain pairs of growing and decaying normal modes. 
We can now relate these normal mode stability properties of the physical system with the fixed point stability of the dynamical system on the phase plane. Equations (\ref{eq:3.3})--(\ref{eq:3.5}) indicate that for 
$|{\mu / 2}| < 1$, we have $\mathcal{D}  = -4\sin{\epsilon^*}$, $\lambda_1^J = \lambda_2^J =  \mathcal{D}^*/2 = -2\lambda_r^{NM}$. Hence, both of the fixed points are star nodes, where the physical growing normal mode is a dynamical sink and the decaying mode is a source in the phase-plane. This apparent contradiction actually makes sense when recalling that the physical solution is the superposition of the two normal modes in Eq.  \eqref{eq:2.8}. Therefore, any initial condition which combines projection on the two normal modes will converge in time to the unstable normal mode as the stable mode decays with time.

\subsubsection{The case of $\, |{\mu \over 2}| > 1$}

When $\mu$ exceeds the absolute value of $2$, 
the ratio between the difference between the waves' frequencies (in the absence of interaction) and the mean interaction coefficient becomes too large to allow modal growth. When $\mu > 2$ the shear is too strong so the waves must be in phase $(\epsilon^* = 0)$ to fully help each other to counter-propagate against the shear. In contrast, When $\mu < -2$ the shear is too weak so the waves must be anti-phased $(\epsilon^* = \pi)$ to fully hinder each others' counter-propagation rate. In both cases Eq.  (\ref{eq:1.1}a) indicates that amplitude growth is prevented.  Furthermore as $|\mu| > 2$, the amplitude symmetry between the waves is broken $(\chi^* \neq 1)$, since  $(\chi +\chi^{-1})^* > 2$  (c.f. Eq.  (\ref{eq:3.1}b)), where the wave with the larger amplitude affects the other more than vice-versa. 

Hence for $\mu > 2$, and when $\chi>1$, the upper wave, with the larger amplitude,  will successfully help the lower one to counter-propagate against its mean flow to end up with a positive frequency relative to $\overline{\omega}$ (rightward propagation). On the other hand, the help provided by  the lower wave (with the smaller amplitude) to the upper one is less effective and as a result, the ability of the upper wave to counter-propagate against the rightward mean flow becomes smaller. As a result the upper wave is ending up as well with a positive frequency relative to $\overline{\omega}$. This is how phase locking is achieved for such neutral modes. Reversing the argument when $\chi<1$ we end up with leftward phase-locked propagation. Following the same logic for $\mu < 2$, and recall that now the waves hinder each others' propagation, we end up with modal leftward propagation for $\chi>1$ and rightward for $\chi<1$. These corresponding compromised frequencies, relative to $\overline{\omega}$, are obtained from the modal eigenvalues Eq.  \eqref{eq:2.11}, where  $\omega^* = -\lambda^{NM}_i = \pm \sqrt{\left ({\mu \over 2}\right )^2 -1}$.

\begin{figure}
\centering\includegraphics[width=0.7\linewidth]{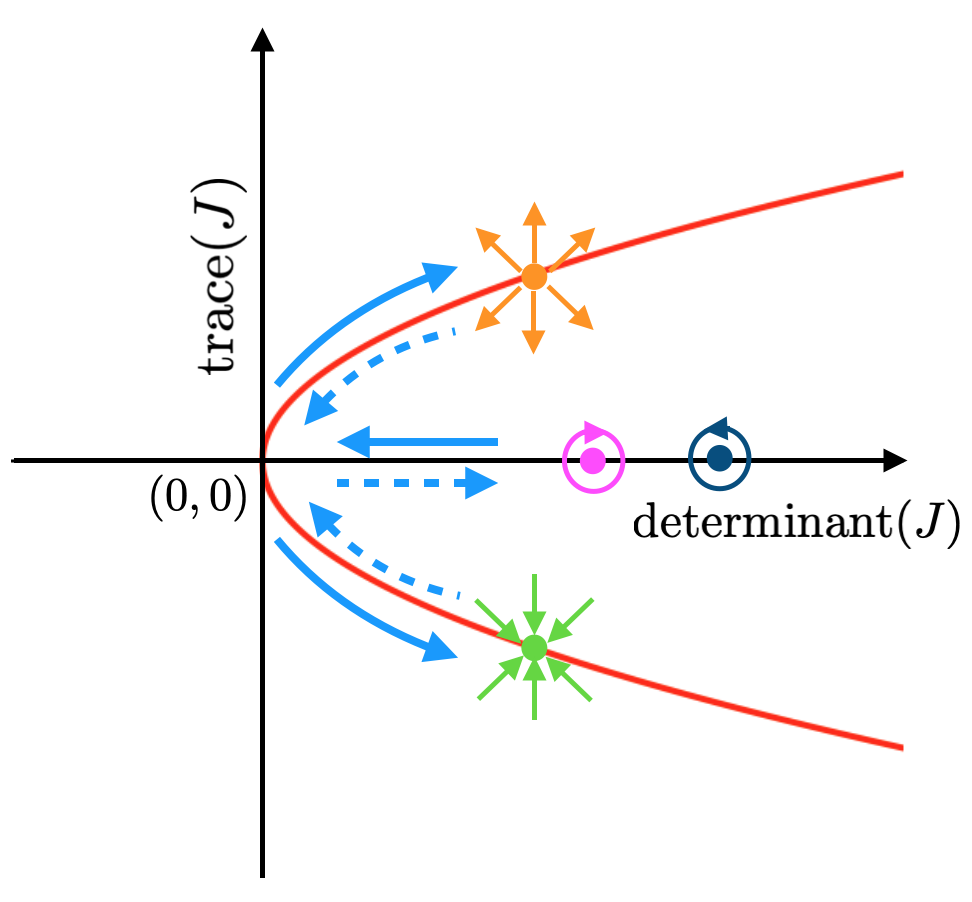}
\caption{Trace-determinant diagram of the Jacobian matrix evaluated at the fixed points. Two neutral center fixed points, respectively shown by dark blue (positive circulation) and magenta (negative circulation)  circles, approach towards the origin along the determinant axis (path shown by the blue solid arrow), as  the bifurcation parameter, $\mu$, is decreased from a high absolute value to $2$ (c.f.\ Fig. \ref{fig:lambda_mu}). A new type of bifurcation occurs at $|\mu|=2$; as $|\mu|$ is further decreased,  a source (filled orange circle with outward arrows) and sink (filled green circle with inward arrows)  fixed points are born. These fixed points lie on the trace-determinant parabola. The exactly  opposite behavior happens when $|\mu|$ is decreased beyond $2$, along the path indicated by the dashed blue arrows.   }
\label{fig:tr_det}
\end{figure}

Since in this regime the vorticity amplitudes ($Q_{1,2}$) of the waves remain small, these neutral modes have generally little relevance for shear instability. Nevertheless, they are interesting from the dynamical perspective. The vertical line, $X = \mu/2$, in the phase plane is an attractor for $Y>0$ and a separatrix for $Y<0$. When $|\mu/2| < 1$, the fixed points are located on the intersection between this vertical line and the unit circle, where the stable star node sits on the attracting side and the unstable one on the separating side. When $|\mu/2| > 1$ the two fixed points sits on the $X$ axis in equal distances from the two sides of $X = \mu/2$, where one point is inside the unit circle and the other is outside of it. As is evident from Equations \eqref{eq:3.3}, \eqref{eq:3.5} and \eqref{eq:3.7}, these points are counter-rotating center points ($\lambda^{J}_{1,2}=\pm\ii |{\chi^{*}}-1/\chi^{*}|$).  Hence, when $|\mu/2| = 1$, the normal form of Eq.  \eqref{eq:2.6} exhibits bifurcation from a pair of a stable and unstable star nodes to two counter-rotating neutral center fixed points; see Fig.\ \ref{fig:tr_det}. We are not familiar with other examples of such type of bifurcation.  

It is interesting to understand the dynamics of these center modes. Perturbing the phase difference $\epsilon$ from the fully helping or the fully hindering setup immediately yields  either a small growth or decay of the wave amplitudes. Since their amplitudes are not even, this growth or decay is more pronounced on the wave with the smaller amplitude, because the wave with the larger amplitude affects the one with the smaller amplitude more efficiently. Thus the amplitude ratio tends to return to its unperturbed value. Similarly,  initially changing the amplitude ratio will unlock the waves and as a result, it changes the amplitude ratio which will in turn act to restore the phase to its neutral position. Hence, near the fixed points, the system  exhibits simple harmonic oscillations. 
To see this mathematically let us write Eq.  \eqref{eq:J3.6} in terms of
$(\chi, \epsilon)$: 
\begin{equation}
\frac{d}{d\tau}\left[\begin{array}{c}
\delta\chi\\
\delta\epsilon
\end{array}\right]=\left[\begin{array}{cc}
-2\chi\sin\epsilon & \left(1-\chi^{2}\right)\cos\epsilon\\
\left(1-\chi^{-2}\right)\cos\epsilon & -\left(\chi+\chi^{-1}\right)\sin\epsilon
\end{array}\right]^*\left[\begin{array}{c}
\delta\chi\\
\delta\epsilon
\end{array}\right].
\end{equation}
In the vicinity of the neutral fixed points we obtain:
\begin{subequations}
 \begin{align}
\frac{d}{d\tau}\delta\chi&=\pm (1-\chi^{2})^* \delta \epsilon,\\
\frac{d}{d\tau}\delta \epsilon &= \pm (1-\chi^{-2})^*\delta\chi,
\end{align}
\label{eq:3.13}
\end{subequations}
where the plus (minus) sign corresponds to the case where $\mu$ is larger (smaller) than $2$. For either cases Eq.  \eqref{eq:3.13} yields:
\begin{equation}
{1\over \delta\chi}\frac{d^2}{d\tau^2}\delta\chi = 
{1\over \delta\epsilon}\frac{d^2}{d\tau^2}\delta\epsilon =
-[(\chi-\chi^{-1})^2]^* = -(\lambda_i^J)^2,
\end{equation}
where $|\lambda_i^J|$ is the oscillation frequency, as expected.

\subsection{A 3D conservative phase space}

The four Equation--set (\ref{eq:1.1}), or equivalently the Hamiltonian system Eq.  \eqref{eq:2.2} of the two waves' action-angle conjugate pairs, were mapped in the previous sub-section into the two Equation--set \eqref{eq:3.1}, or equivalently into Eq.  \eqref{eq:3.2}. Hence, a four degrees of freedom Hamiltonian system was mapped into a degenerated non-Hamiltonian system with only two degrees of freedom.  In the former, the fixed points are centers, representing solutions in which the two waves are both neutral and stationary. In the latter, area is not conserved in the phase plane, which allows sink and source fixed points to represent modal growth and decay, respectively. The reason that such mapping is at all possible  is that the nonlinear system Eq.  \eqref{eq:1.1} emanates from the linearised system Eq.  \eqref{eq:2.3}, which can be determined only up to an arbitrary complex scaling factor between $q_1$ and $q_2$. The degeneracy results from the fact that the amplitude of such scaling factor is ``hidden'' inside $\chi$ and its phase inside $\epsilon$.

\begin{figure}
\centering\includegraphics[width=0.9\textwidth]{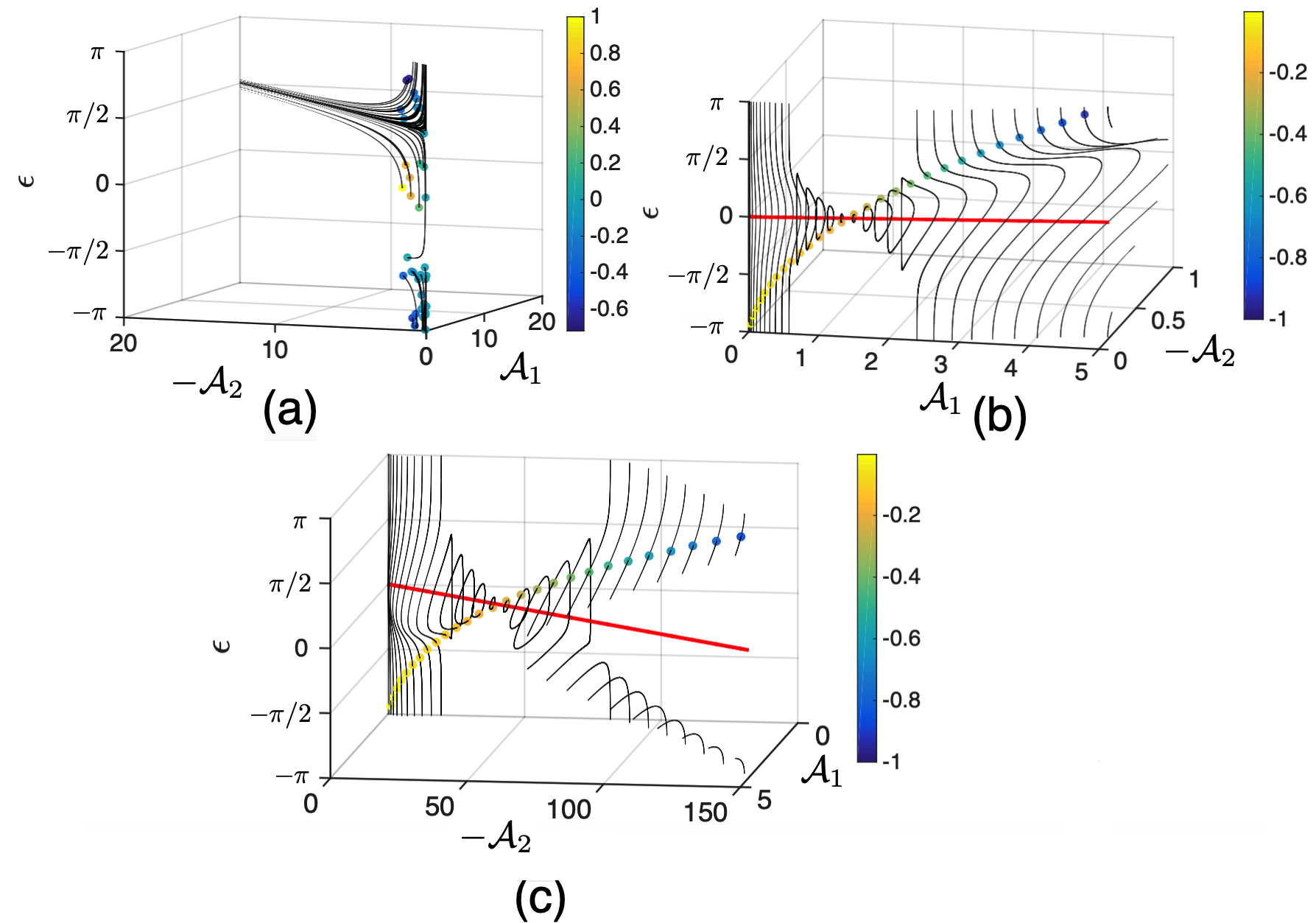}
\caption{Phase portrait showing selected  trajectories in the conservative phase space ${\cal A}_1$--$ {\cal A}_2$ -- $\epsilon$. The Rayleigh instability problem of section \ref{Section 4} has been chosen. (a) Unstable case: $\mu=0$ and initial points located along ${\cal A}_1 = -{\cal A}_2$, (b) Stable case: $\mu=3$ and initial points located along ${\cal A}_1 = -\left[{\mu^2/2}-1 + \mu\sqrt{\left ({\mu/ 2}\right )^2-1}\,\right]{\cal A}_2$, and (c) Stable case: $\mu=3$ and initial points located along ${\cal A}_1 = -\left[{\mu^2 /2}-1 - \mu\sqrt{\left ({\mu / 2}\right )^2-1}\,\right]{\cal A}_2$. The red lines denote  ${\cal A}_1 = -\left[{\mu^2 /2}-1 +\pm \mu\sqrt{\left ({\mu / 2}\right )^2-1}\,\right]{\cal A}_2$ for $\epsilon=0$. Colors show the normalized pseudoenergy, which, being a constant of motion, remains constant along each trajectory.}
\label{fig::A1A2eps}
\end{figure}

In between the (impractical for demonstration) Hamiltonian 4D configuration phase space and the compact non-conservative phase-plane we note that the Hamiltonian ${\cal H}$ is a function of $A_1$, $A_2$, and $\epsilon$, and not separately of $\epsilon_1$ and $\epsilon_2$, see Eq.  (\ref{eq:2.1}a). This allows the construction of a 3D volume preserving phase space out of these variables. The dynamical equations on this phase space are given by Eq.  (\ref{eq:2.2}a) (two equations) and the subtraction of the two equations (${\dot \epsilon}_{1}$ from ${\dot \epsilon}_{2}$) of Eq.  (\ref{eq:2.2}b), yielding:
\begin{equation}\label{eq:3.15}
{\dot \epsilon} = - {\hat \omega} + \ii\sigma\left (\sqrt{{\cal A}_{2}\over {\cal A}_{1}} - \sqrt{{\cal A}_{1}\over {\cal A}_{2}}\right ) \cos{\epsilon} = \der{{\cal H}}{{\cal A}_{1}} - \der{{\cal H}}{{\cal A}_{2}} \, .
\end{equation}
Volume is then conserved in this $A_1$--$A_2$--$\epsilon$ phase space:
$$\der{\dot{\cal A}_1}{{\cal A}_1} + 
\der{\dot{\cal A}_2}{{\cal A}_2} + \der{\dot{\epsilon}}{\epsilon}= 0,$$ as can be easily proved using the facts that: $$\dot{\cal A}_{1,2} = -\der{{\cal H}}{\epsilon_{1,2}}\,\,\,\mathrm{and}\,\,\, \der{\epsilon}{\epsilon_{1}}= 1\,\, \& \,\, \der{\epsilon}{\epsilon_{2}}= -1 .$$
In this phase space the stable and unstable star nodes Eq.  (\ref{eq:3.4}a) of the 2D phase space  are mapped into the lines satisfying 
${\cal A}_1 = -{\cal A}_2$ at the level of $\epsilon^*  = \pm \cos^{-1}\left({\mu / 2}\right)$, see Fig.\ \ref{fig::A1A2eps}a.  The neutral center fixed points of Eqs. (\ref{eq:3.4}b)--(\ref{eq:3.4}c) are obtained for the lines 
$${\cal A}_1 = -\left[{\mu^2 \over 2}-1 \pm \mu\sqrt{\left ({\mu \over 2}\right )^2-1}\,\right]{\cal A}_2,$$ on the surfaces $\epsilon^* = 0$ and $\pi$ respectively; as can be understood from Figs. \ref{fig::A1A2eps}b--\ref{fig::A1A2eps}c. 

\begin{figure}
\centering\includegraphics[width=0.7\linewidth]{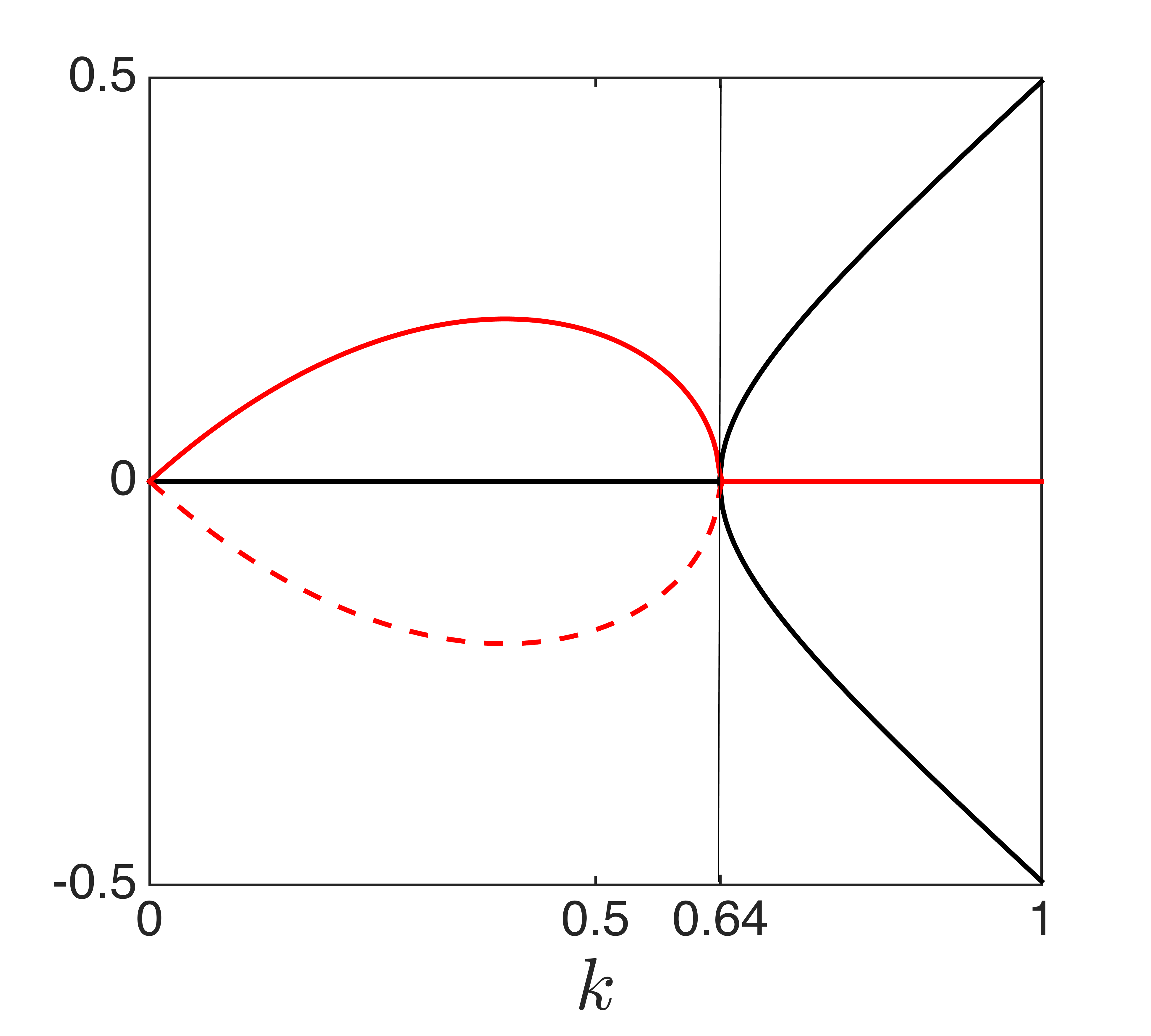}
\caption{Bifurcation diagram for Rayleigh instability. Black line denotes $\lambda_i$, solid  red line denotes $\lambda_r>0$, marking the unstable normal mode, while the dashed red line denotes $\lambda_r<0$, i.e.\ stable normal mode. Bifurcation occurs when $k=0.64$ (shown by the thin vertical line). Direct comparison can be made with Fig.\ \ref{fig:lambda_mu}, noting that here $\mu$ is always greater than $-2$.}
\label{fig::rayleigh_bf}
\end{figure}

\section{Application to the Rayleigh model of shear instability}
\label{Section 4}

We wish to provide a concrete example for the wave interaction mechanism. One of the simplest setups for shear instability has been suggested by Lord Rayleigh in (1880) \cite[]{draz1982} for a piecewise version of the shear profile in Fig.\ \ref{fig:1}:
\begin{equation}
U\left(y\right)=\begin{cases}
\,\,\,\,1 & \,\,\,\,\,\,\,\,\,\,\,\,\,\,\,\,y\geq1\\
\,\,\,\,y & -1\leq y\leq1\\
-1 & \,\,\,\,\,\,\,\,\,\,\,\,\,\,\,y\leq-1.\end{cases}\label{eq:kh2}
\end{equation}
Detailed analysis of the problem in terms of wave interaction can be found in Ref.\, \cite{heif1999} for modal instability, and in Ref.\, \cite{heif2005} for non-modal growth. Here we note that for this piecewise version of shear profile the mean vorticity gradient is concentrated in $y = \pm 1$:
\begin{equation}
\Omega_y=-{U}_{yy} =\delta\left(y-1\right)-\delta\left(y+1\right).
\label{eq:delta_behav}
\end{equation}
thus the two vorticity waves are interfacial so that the perturbation vorticity
$q$ satisfies:
\begin{equation}
q=\left[q_1(k,t)\delta\left(y-1\right)+q_2(k,t)\delta\left(y+1\right)\right]\ee^{\ii k x},
\end{equation}
where $k>0$ denotes the streamwise wavenumber. Analysis of this symmetric setup reveals that $\sigma = \sigma_1 = \sigma_2 =  {\ee^{-2k}/2}$ and ${\hat \omega_1} = -{\hat \omega_2} = k-{1/2}$, yielding ${\overline \omega} =0$ and 
${\mu / 2}=(2k-1)\ee^{2k}$. Hence, the wavenumber is the actual control parameter of the problem which is mapped into the control parameter $\mu$ of the normal form of Eq.  \eqref{eq:2.6}. Since $k$ is positive, ${\mu / 2}>-1$, therefore $k=0$ is the limit of two infinitely fast counter-propagating waves that must be in anti-phase to fully hinder each other's propagation rate in order to remain phase-locked. Furthermore, $k=0.5$ corresponds to $\mu=0$ where the waves are either $\pi/2$ or $-\pi/2$ out of phase for the growing and decaying modes, respectively. Since ${\mu / 2}$ becomes larger than $1$ for $k_c >0.64$, it therefore implies that for  wavenumbers larger than $k_c$, all normal modes are neural.  In this scenario, the waves are in phase to fully help each other to counter-propagate against the shear, however the wave amplitudes are not even. The bifurcation diagram for Rayleigh's shear instability is given in Fig.\ \ref{fig::rayleigh_bf}.  
 
\section{Multi--wave interactions}
\label{Section 5}

The interaction between the two vorticity waves, described in  Eq.  \eqref{eq:1.1}, can be generalized straightforwardly to the case of $N$ number of interacting waves, as is illustrated schematically in Fig.\, \ref{fig:Multi waves},
and explained in the Appendix of Ref. \cite{heifetz2009canonical} for Rossby waves and in Ref. \cite{heifetz2018generalized} for gravity waves. 
The cross-stream velocity field at every level is now composed of the in-situ velocity field, induced by the wave located at that level (indicated hereafter by the index $i$, and the contributions of the far field velocity induced by all of the other remote waves, indicated generally by the index $j$). Naturally, the magnitude of the induced velocity decreases with the distance according to the evanescent structure of the Green function, which translates the vorticity source to the far field velocity field it induces. This structure is determined by the details of the problem setup (c.f. different examples in the Appendix of Ref. \cite{heif2005}) however it only depends on the the cross-stream distance between waves $i$ and $j$. In the following formulation $G_{ij}$ represents the Green function induced by a remote wave $j$ on an in-situ wave $i$. Since only the distance between the two waves matters for $G$, it is symmetric, i.e., $G_{ij}=G_{ji}$.
Generally we may expect that adjacent pairs of vorticity waves will affect each other more pronouncedly  than remote pairs. Nonetheless, a remote vorticity wave with a large amplitude $Q_j$ may affect a distant wave more strongly than a closer neighbour wave with a smaller amplitude. 

\begin{figure}
\centering\includegraphics[width=0.9\textwidth]{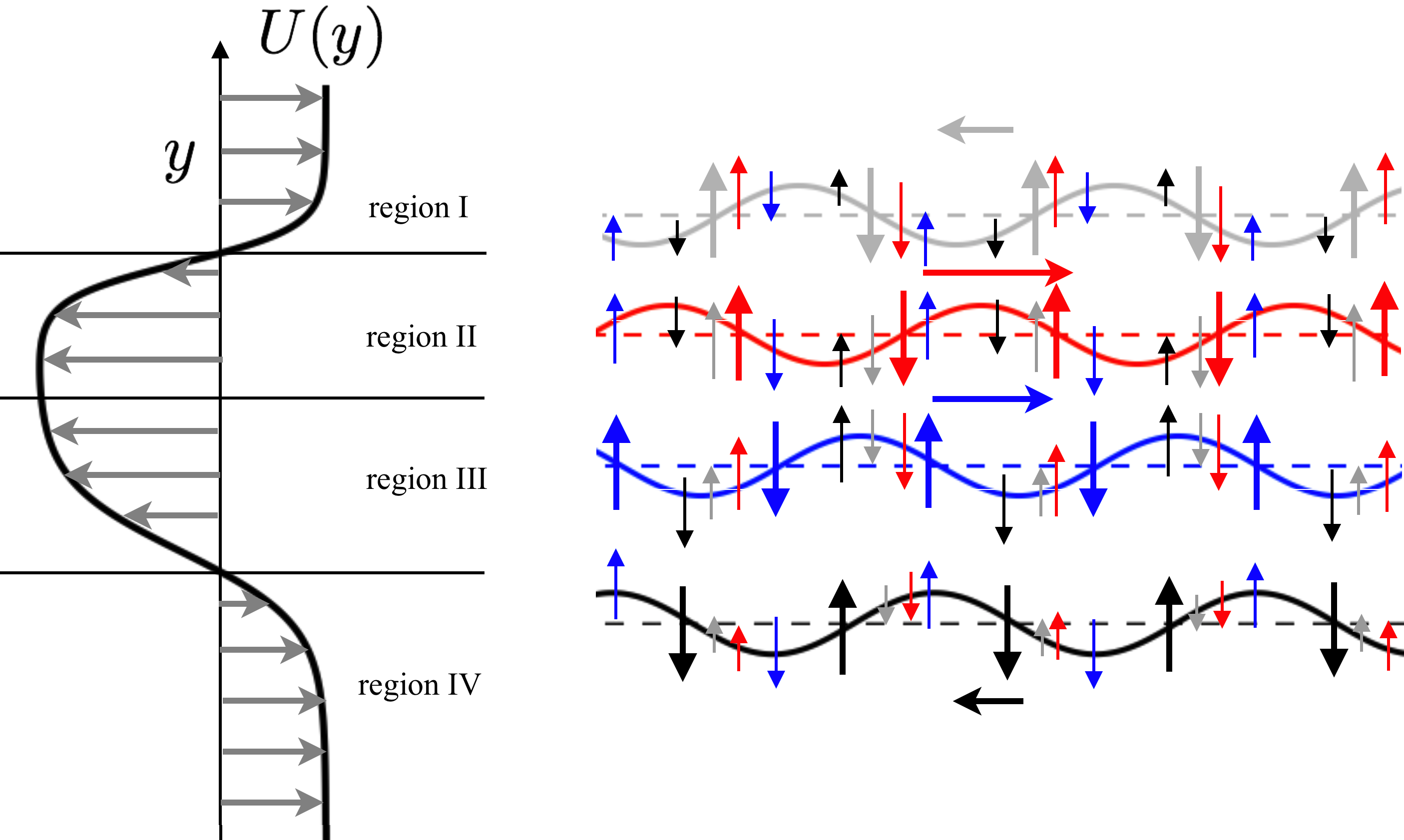}
\caption{Schematic of a general shear layer, the complex instability dynamics of which can be understood using the  minimal model of $N$ ($=4$ in this case) interacting vorticity waves. The color convention is same as that of Fig.\ \ref{fig:1}.}
\label{fig:Multi waves}
\end{figure}

Furthermore, as illustrated in Figs. \ref{fig:1} and \ref{fig:Multi waves} the cross-stream velocity field acts directly on the wave displacement. Thus, if the displacement and the vorticity wave anomalies are in phase (like in wave `$2$' in Fig. \ref{fig:1}) a positive far field cross-stream velocity, acting to amplify the wave displacement, is also amplifying the positive vorticity anomaly. In contrast, when the wave displacement and vorticity anomalies are in anti-phase (as in wave `$1$' in Fig. \ref{fig:1}) such far field velocity will increase the negative value of the vorticity anomaly. The amount by which an induced velocity increases the vorticity amplitude of an in-situ wave depends on the restoring mechanism of the wave itself and is generally different for Rossby, gravity, capillary or Alfven waves. As we are interested in the prototype of the interaction we therefore indicate this factor by $\alpha_i$, which is positive when the displacement and the vorticity wave anomalies of wave $i$ are in anti-phase and negative when they are in phase. Denote $\epsilon_{ij} \equiv \epsilon_{i}-\epsilon_{j}$, the generalization of Eq.  \eqref{eq:1.1} to $N$ interaction waves read: 
\begin{subequations}\label{eq:Multi}
\begin{align}
{\dot Q}_{i}& = \alpha_i\sum_{j=1}^{N} Q_{j}G_{ij}\sin{\epsilon_{ij}},   \\
{\dot \epsilon}_{i} & = - {\hat \omega}_{i} + {\alpha_i\over Q_{i}} \sum_{j=1, j\neq i}^{N}  Q_{j}G_{ij} \cos{\epsilon_{ij}}.
\end{align}
\end{subequations}
Thus, from this wave interaction perspective modal phase locking is achieved when a configuration is set to synchronize all waves to propagate with the same frequency: $-\omega^* = \lambda_i^{NM} ={\dot \epsilon}_{1} = {\dot \epsilon}_{2}=...={\dot \epsilon}_{N}$, and to exhibit the same growth rate $\lambda_r^{NM} = {{\dot Q}_{1}/ Q_1}= {{\dot Q}_{2}/ Q_2}=...={{\dot Q}_{N}/ Q_N}$.

The wave action and the pseudo-energy conservation laws for Eq.  \eqref{eq:Multi}: 
\begin{equation}\label{eq:Multi PM PE}
{\cal A}  = \sum_{i=1}^{N} {\cal A}_i  = \sum_{i=1}^{N} {Q_i^2\over 2\alpha_i}\, ; \qquad
{\cal H}  =\sum_{i=1}^{N} {\cal A}_i {\dot \epsilon}_{i}\, ,
\end{equation}
satisfy then the Hamilton equations:

\begin{equation}\label{eq:Multi Hamilton}
{\dot {\cal A}}_{i} = -\der{{\cal H}}{\epsilon_{i}}\, ; \qquad  {\dot \epsilon}_{i} =   \der{{\cal H}}{{\cal A}_{i}}.
\end{equation}

\section{Conclusions}
\label{Section 6} 

In this paper we have examined a simple, nonlinear, autonomous dynamical system which describes some central aspects of 2D shear instability. The building blocks of the system are interacting counter propagating vorticity waves. Instability is achieved when the waves are synchronized to propagate in a phase-locked configuration which allows mutual amplification, i.e., resonance. The dynamics  originates from the linearised vorticity equation for shear flows and therefore valid only for small wave amplitudes. Hence, triad interactions between waves with different wavenumbers are excluded. Nonetheless, the interaction between distant waves of the same wavenumber across the shear is nonlinear. Mathematically, this nonlinear representation of linearised dynamics results from introducing the perturbation field in its  polar form, i.e., in terms of amplitude and phase. However, this apparent additional complication provides  a clear mechanistic interpretation for the instability mechanism. A somewhat similar example in which some aspects of linear dynamics  become clearer when introduced in its nonlinear polar form is the Madelung equation \cite{madelung1927quantentheorie}, which converts the linear Schrodinger equation into a fluid dynamic, Euler-like equation \cite{heifetz2015toward}. 

The wave interaction equations are conservative. In the generalized configuration space $\mathbb{R}^{2N}$ (where $N$ is the number of the interacting waves) the pseudo-energy serves as the Hamiltonian, the square of the wave amplitudes are proportional to  their wave-action, and their phases can be considered as angles. It is then straightforward to show that the system satisfies a generalized action-angle Hamilton equations where the total action of all the waves is conserved, however the action of each  individual wave may change (unlike the classical action-angle formalism) due to the action-at-a-distance interaction between the waves.  

Since the wave interaction equations emanate from the linearised dynamics, it is determined up to an arbitrary complex scaling factor between the waves. As a result, the essence of the wave interaction dynamics can be described in a reduced non-conservative phase space with only $N$ degrees of freedom. In this `reduced' phase space, unstable normal modes of the linearised system are represented by stable star fixed points, and the stable normal modes by unstable star fixed points. This apparent contradiction actually makes sense since in the linearised system, a perturbation solution is a superposition of the unstable (growing) and stable (decaying) normal modes. Hence, as time evolves, the perturbation will be biased towards the unstable normal mode solution, diverging away from the unstable fixed point (in the reduced phase space) and converging towards the stable fixed point. 
If the shear is either too strong or too weak to allow resonance, synchronization between the waves is still possible. In these scenarios, the waves are phase-locked to propagate in concert with the same frequency, however the amplitude of the waves does not change due to the interaction between the waves. In the linearised description, these configurations describe neutral normal modes, whereas in the reduced phase space, these are neutral central fixed points.   

Furthermore, for the two--wave interaction problem, the dynamics in the reduced phase-space can be succinctly expressed as a complex normal form equation for the normalized perturbation vorticity ratio between the waves.
It is a non-homogeneous equation (since the shear acts as an exterior forcing) which includes a single control parameter. The latter is the ratio between the differences between the  waves' frequencies in the absence of interaction and the interaction coefficient. This makes sense as the waves generally tend to propagate in opposite directions in the absence of interaction, whereas the mutual interaction tends to keep them together.      
This normal form exhibits bifurcation where annihilation of a pair of stable and unstable star nodes yields the emergence of two  neutral center fixed points of opposite circulations. To the best of our knowledge, this is a new type of bifurcation.

The two-wave and the general $N$-wave interaction dynamics described here can be regarded as well as a novel model for synchronization. Each agent (wave) in isolation acts to resist (counter-propagate against) a local external forcing (the mean flow shear); some agents counter-propagate more efficiently than  others. Hence, alignment (phase locking) is achieved only through overall collaboration (far field interaction) between the agents. The `too efficient' agents should be hindered by the overall interaction whereas the `less efficient' ones should be helped. This dynamics shares some similarities with the Kuramoto  model, however it differs from the latter since here each agent does not try to adjust its frequency to the other to obtain synchronization. In contrast, when the waves are in phase, they act to increase their phase difference.

A further novel aspect of our model is that alignment can lead to mutual amplification of the agents' amplitudes (modal instability). Hence, eventually the agents will be strong enough not only to interact between themselves, but also to alter their `environmental averages' conditions  \cite{motsch2014heterophilious}, i.e.\, the mean flow. This is one of the central initial mechanisms by which 2D laminar shear flows are transformed into a turbulent state. A straightforward generalization of such wave-mean flow interaction is currently being  studied by the authors. 

The interaction described here is a type of long-range interaction. As illustrated in Fig. \ref{fig:Multi waves} the instantaneous interaction between each pair of distant waves is not affected by the waves sandwiched in between. However, in reality, for finite values of the Reynolds number, viscosity may play a vital role in the dynamics and should be represented by a short-range interaction \cite{shvydkoy2018topological} between the agents. Another additional piece of reality is that shear flows in nature are generally  continuously exposed to some level of noise (both exterior and interior due to triad interaction processes), which affects the mean flow and the waves.  Diffusive \cite{shvydkoy2018topological} and stochastic  \cite{during2009boltzmann} processes have already been implemented in collective dynamics in different contexts, and in nonlinear dynamical systems. While wave--mean flow models already exist \cite{pedlosky:1990}, there remains a considerable scope towards understanding it from the wave interactions perspective. In near future, the authors aim to provide a more realistic description of counter-propagating wave-mean flow interactions as a forced-dissipative system.

\begin{acknowledgments}
A.G. would like to acknowledge the funding support from the Alexander von Humboldt foundation.
\end{acknowledgments}

\bibliography{apssamp}

 \newcommand{\noop}[1]{}
\begin{thebibliography}{22}%
\makeatletter
\providecommand \@ifxundefined [1]{%
 \@ifx{#1\undefined}
}%
\providecommand \@ifnum [1]{%
 \ifnum #1\expandafter \@firstoftwo
 \else \expandafter \@secondoftwo
 \fi
}%
\providecommand \@ifx [1]{%
 \ifx #1\expandafter \@firstoftwo
 \else \expandafter \@secondoftwo
 \fi
}%
\providecommand \natexlab [1]{#1}%
\providecommand \enquote  [1]{``#1''}%
\providecommand \bibnamefont  [1]{#1}%
\providecommand \bibfnamefont [1]{#1}%
\providecommand \citenamefont [1]{#1}%
\providecommand \href@noop [0]{\@secondoftwo}%
\providecommand \href [0]{\begingroup \@sanitize@url \@href}%
\providecommand \@href[1]{\@@startlink{#1}\@@href}%
\providecommand \@@href[1]{\endgroup#1\@@endlink}%
\providecommand \@sanitize@url [0]{\catcode `\\12\catcode `\$12\catcode
  `\&12\catcode `\#12\catcode `\^12\catcode `\_12\catcode `\%12\relax}%
\providecommand \@@startlink[1]{}%
\providecommand \@@endlink[0]{}%
\providecommand \url  [0]{\begingroup\@sanitize@url \@url }%
\providecommand \@url [1]{\endgroup\@href {#1}{\urlprefix }}%
\providecommand \urlprefix  [0]{URL }%
\providecommand \Eprint [0]{\href }%
\providecommand \doibase [0]{https://doi.org/}%
\providecommand \selectlanguage [0]{\@gobble}%
\providecommand \bibinfo  [0]{\@secondoftwo}%
\providecommand \bibfield  [0]{\@secondoftwo}%
\providecommand \translation [1]{[#1]}%
\providecommand \BibitemOpen [0]{}%
\providecommand \bibitemStop [0]{}%
\providecommand \bibitemNoStop [0]{.\EOS\space}%
\providecommand \EOS [0]{\spacefactor3000\relax}%
\providecommand \BibitemShut  [1]{\csname bibitem#1\endcsname}%
\let\auto@bib@innerbib\@empty
\bibitem [{\citenamefont {Hoskins}\ \emph {et~al.}(1985)\citenamefont
  {Hoskins}, \citenamefont {McIntyre},\ and\ \citenamefont
  {Robertson}}]{hosk1985}%
  \BibitemOpen
  \bibfield  {author} {\bibinfo {author} {\bibfnamefont {B.~J.}\ \bibnamefont
  {Hoskins}}, \bibinfo {author} {\bibfnamefont {M.~E.}\ \bibnamefont
  {McIntyre}},\ and\ \bibinfo {author} {\bibfnamefont {A.~W.}\ \bibnamefont
  {Robertson}},\ }\bibfield  {title} {\bibinfo {title} {On the use and
  significance of isentropic potential vorticity maps},\ }\href@noop {}
  {\bibfield  {journal} {\bibinfo  {journal} {Q. J. Roy. Meteor. Soc.}\
  }\textbf {\bibinfo {volume} {111}},\ \bibinfo {pages} {877} (\bibinfo {year}
  {1985})}\BibitemShut {NoStop}%
\bibitem [{\citenamefont {Caulfield}(1994)}]{caul1994}%
  \BibitemOpen
  \bibfield  {author} {\bibinfo {author} {\bibfnamefont {C.~P.}\ \bibnamefont
  {Caulfield}},\ }\bibfield  {title} {\bibinfo {title} {Multiple linear
  instability of layered stratified shear flow},\ }\href@noop {} {\bibfield
  {journal} {\bibinfo  {journal} {J. Fluid Mech.}\ }\textbf {\bibinfo {volume}
  {258}},\ \bibinfo {pages} {255} (\bibinfo {year} {1994})}\BibitemShut
  {NoStop}%
\bibitem [{\citenamefont {Heifetz}\ \emph {et~al.}(1999)\citenamefont
  {Heifetz}, \citenamefont {Bishop},\ and\ \citenamefont {Alpert}}]{heif1999}%
  \BibitemOpen
  \bibfield  {author} {\bibinfo {author} {\bibfnamefont {E.}~\bibnamefont
  {Heifetz}}, \bibinfo {author} {\bibfnamefont {C.~H.}\ \bibnamefont
  {Bishop}},\ and\ \bibinfo {author} {\bibfnamefont {P.}~\bibnamefont
  {Alpert}},\ }\bibfield  {title} {\bibinfo {title} {Counter-propagating
  {R}ossby waves in the barotropic {R}ayleigh model of shear instability},\
  }\href@noop {} {\bibfield  {journal} {\bibinfo  {journal} {Q. J. R. Meteorol.
  Soc.}\ }\textbf {\bibinfo {volume} {125}},\ \bibinfo {pages} {2835} (\bibinfo
  {year} {1999})}\BibitemShut {NoStop}%
\bibitem [{\citenamefont {Heifetz}\ and\ \citenamefont
  {Methven}(2005)}]{heif2005}%
  \BibitemOpen
  \bibfield  {author} {\bibinfo {author} {\bibfnamefont {E.}~\bibnamefont
  {Heifetz}}\ and\ \bibinfo {author} {\bibfnamefont {J.}~\bibnamefont
  {Methven}},\ }\bibfield  {title} {\bibinfo {title} {Relating optimal growth
  to counterpropagating {R}ossby waves in shear instability},\ }\href@noop {}
  {\bibfield  {journal} {\bibinfo  {journal} {Phys. Fluids}\ }\textbf {\bibinfo
  {volume} {17}},\ \bibinfo {eid} {064107} (\bibinfo {year}
  {2005})}\BibitemShut {NoStop}%
\bibitem [{\citenamefont {Carpenter}\ \emph {et~al.}(2013)\citenamefont
  {Carpenter}, \citenamefont {Tedford}, \citenamefont {Heifetz},\ and\
  \citenamefont {Lawrence}}]{carp2012}%
  \BibitemOpen
  \bibfield  {author} {\bibinfo {author} {\bibfnamefont {J.~R.}\ \bibnamefont
  {Carpenter}}, \bibinfo {author} {\bibfnamefont {E.~W.}\ \bibnamefont
  {Tedford}}, \bibinfo {author} {\bibfnamefont {E.}~\bibnamefont {Heifetz}},\
  and\ \bibinfo {author} {\bibfnamefont {G.~A.}\ \bibnamefont {Lawrence}},\
  }\bibfield  {title} {\bibinfo {title} {Instability in stratified shear flow:
  Review of a physical interpretation based on interacting waves},\ }\href@noop
  {} {\bibfield  {journal} {\bibinfo  {journal} {Appl. Mech. Rev.}\ }\textbf
  {\bibinfo {volume} {64}},\ \bibinfo {pages} {060801} (\bibinfo {year}
  {2013})}\BibitemShut {NoStop}%
\bibitem [{\citenamefont {Guha}\ and\ \citenamefont
  {Lawrence}(2014)}]{guha2014}%
  \BibitemOpen
  \bibfield  {author} {\bibinfo {author} {\bibfnamefont {A.}~\bibnamefont
  {Guha}}\ and\ \bibinfo {author} {\bibfnamefont {G.~A.}\ \bibnamefont
  {Lawrence}},\ }\bibfield  {title} {\bibinfo {title} {A wave interaction
  approach to studying non-modal homogeneous and stratified shear
  instabilities},\ }\href@noop {} {\bibfield  {journal} {\bibinfo  {journal}
  {J. Fluid Mech.}\ }\textbf {\bibinfo {volume} {755}},\ \bibinfo {pages} {336}
  (\bibinfo {year} {2014})}\BibitemShut {NoStop}%
\bibitem [{\citenamefont {Harnik}\ \emph {et~al.}(2008)\citenamefont {Harnik},
  \citenamefont {Heifetz}, \citenamefont {Umurhan},\ and\ \citenamefont
  {Lott}}]{harnik2008}%
  \BibitemOpen
  \bibfield  {author} {\bibinfo {author} {\bibfnamefont {N.}~\bibnamefont
  {Harnik}}, \bibinfo {author} {\bibfnamefont {E.}~\bibnamefont {Heifetz}},
  \bibinfo {author} {\bibfnamefont {O.~M.}\ \bibnamefont {Umurhan}},\ and\
  \bibinfo {author} {\bibfnamefont {F.}~\bibnamefont {Lott}},\ }\bibfield
  {title} {\bibinfo {title} {A buoyancy-vorticity wave interaction approach to
  stratified shear flow},\ }\href@noop {} {\bibfield  {journal} {\bibinfo
  {journal} {J. Atmos. Sci.}\ }\textbf {\bibinfo {volume} {65}},\ \bibinfo
  {pages} {2615} (\bibinfo {year} {2008})}\BibitemShut {NoStop}%
\bibitem [{\citenamefont {Biancofiore}\ \emph {et~al.}(2015)\citenamefont
  {Biancofiore}, \citenamefont {Gallaire},\ and\ \citenamefont
  {Heifetz}}]{biancofiore2015}%
  \BibitemOpen
  \bibfield  {author} {\bibinfo {author} {\bibfnamefont {L.}~\bibnamefont
  {Biancofiore}}, \bibinfo {author} {\bibfnamefont {F.}~\bibnamefont
  {Gallaire}},\ and\ \bibinfo {author} {\bibfnamefont {E.}~\bibnamefont
  {Heifetz}},\ }\bibfield  {title} {\bibinfo {title} {Interaction between
  counterpropagating {R}ossby waves and capillary waves in planar shear
  flows},\ }\href@noop {} {\bibfield  {journal} {\bibinfo  {journal} {Phys.
  Fluids}\ }\textbf {\bibinfo {volume} {27}},\ \bibinfo {pages} {044104}
  (\bibinfo {year} {2015})}\BibitemShut {NoStop}%
\bibitem [{\citenamefont {Heifetz}\ \emph {et~al.}(2015)\citenamefont
  {Heifetz}, \citenamefont {Mak}, \citenamefont {Nycander},\ and\ \citenamefont
  {Umurhan}}]{heifetzalfven2015}%
  \BibitemOpen
  \bibfield  {author} {\bibinfo {author} {\bibfnamefont {E.}~\bibnamefont
  {Heifetz}}, \bibinfo {author} {\bibfnamefont {J.}~\bibnamefont {Mak}},
  \bibinfo {author} {\bibfnamefont {J.}~\bibnamefont {Nycander}},\ and\
  \bibinfo {author} {\bibfnamefont {O.~M.}\ \bibnamefont {Umurhan}},\
  }\bibfield  {title} {\bibinfo {title} {Interacting vorticity waves as an
  instability mechanism for magnetohydrodynamic shear instabilities},\
  }\href@noop {} {\bibfield  {journal} {\bibinfo  {journal} {J. Fluid Mech.}\
  }\textbf {\bibinfo {volume} {767}},\ \bibinfo {pages} {199} (\bibinfo {year}
  {2015})}\BibitemShut {NoStop}%
\bibitem [{Note1()}]{Note1}%
  \BibitemOpen
  \bibinfo {note} {Note that in Fig. \ref {fig:2}, the displacement fields have
  been plotted. In Fig. \ref {fig:2}(a) the displacement fields are anti-phased
  however the vorticity fields are in phase.}\BibitemShut {Stop}%
\bibitem [{\citenamefont {Heifetz}\ \emph {et~al.}(2004)\citenamefont
  {Heifetz}, \citenamefont {Bishop}, \citenamefont {Hoskins},\ and\
  \citenamefont {Methven}}]{heifetz2004counter}%
  \BibitemOpen
  \bibfield  {author} {\bibinfo {author} {\bibfnamefont {E.}~\bibnamefont
  {Heifetz}}, \bibinfo {author} {\bibfnamefont {C.}~\bibnamefont {Bishop}},
  \bibinfo {author} {\bibfnamefont {B.}~\bibnamefont {Hoskins}},\ and\ \bibinfo
  {author} {\bibfnamefont {J.}~\bibnamefont {Methven}},\ }\bibfield  {title}
  {\bibinfo {title} {The counter-propagating {R}ossby-wave perspective on
  baroclinic instability. {I}: Mathematical basis},\ }\href@noop {} {\bibfield
  {journal} {\bibinfo  {journal} {Q. J. Roy. Meteor. Soc.}\ }\textbf {\bibinfo
  {volume} {130}},\ \bibinfo {pages} {211} (\bibinfo {year}
  {2004})}\BibitemShut {NoStop}%
\bibitem [{\citenamefont {Heifetz}\ and\ \citenamefont
  {Guha}(2018)}]{heifetz2018generalized}%
  \BibitemOpen
  \bibfield  {author} {\bibinfo {author} {\bibfnamefont {E.}~\bibnamefont
  {Heifetz}}\ and\ \bibinfo {author} {\bibfnamefont {A.}~\bibnamefont {Guha}},\
  }\bibfield  {title} {\bibinfo {title} {A generalized action-angle
  representation of wave interaction in stratified shear flows},\ }\href@noop
  {} {\bibfield  {journal} {\bibinfo  {journal} {Journal of Fluid Mechanics}\
  }\textbf {\bibinfo {volume} {834}},\ \bibinfo {pages} {220} (\bibinfo {year}
  {2018})}\BibitemShut {NoStop}%
\bibitem [{Note2()}]{Note2}%
  \BibitemOpen
  \bibinfo {note} {In order to see that substitute Eq. \protect \textup {\hbox
  {\mathsurround \z@ \protect \normalfont (\ignorespaces \ref {eq:2.10}\unskip
  \@@italiccorr )}} in Eqs. (\ref {eq:2.1}a)--(\ref {eq:2.1}b) to obtain :
  ${\protect \cal H} = \left [-{\protect \mathaccentV {hat}05E\omega }_1
  {P_1^2\over 2\sigma _1} + {\protect \mathaccentV {hat}05E\omega
  }_2{P_2^2\over 2\sigma _2}+{\mu \over 2} P_{1}P_{2} \right ] e^{2\lambda _r
  t}\protect \tmspace +\thinmuskip {.1667em}$; \par ${\protect \cal A} = \left
  ({P_1^2\over 2\sigma _1} -{P_2^2\over 2\sigma _2}\right )e^{2\lambda _r t}$.
  Therefore both $({\protect \cal H}, {\protect \cal A})$ must vanish to remain
  constant when $\lambda _r \not =0$.}\BibitemShut {Stop}%
\bibitem [{\citenamefont {Pikovsky}\ \emph {et~al.}(2001)\citenamefont
  {Pikovsky}, \citenamefont {Rosenblum},\ and\ \citenamefont {Kurths}}]{PIK01}%
  \BibitemOpen
  \bibfield  {author} {\bibinfo {author} {\bibfnamefont {A.}~\bibnamefont
  {Pikovsky}}, \bibinfo {author} {\bibfnamefont {M.~G.}\ \bibnamefont
  {Rosenblum}},\ and\ \bibinfo {author} {\bibfnamefont {J.}~\bibnamefont
  {Kurths}},\ }\href@noop {} {\emph {\bibinfo {title} {Synchronization, A
  Universal Concept in Nonlinear Sciences}}}\ (\bibinfo  {publisher} {Cambridge
  University Press},\ \bibinfo {year} {2001})\BibitemShut {NoStop}%
\bibitem [{\citenamefont {Drazin}\ and\ \citenamefont {Reid}(2004)}]{draz1982}%
  \BibitemOpen
  \bibfield  {author} {\bibinfo {author} {\bibfnamefont {P.~G.}\ \bibnamefont
  {Drazin}}\ and\ \bibinfo {author} {\bibfnamefont {W.~H.}\ \bibnamefont
  {Reid}},\ }\href@noop {} {\emph {\bibinfo {title} {{H}ydrodynamic
  {S}tability}}},\ \bibinfo {edition} {2nd}\ ed.\ (\bibinfo  {publisher}
  {Cambridge University Press},\ \bibinfo {year} {2004})\BibitemShut {NoStop}%
\bibitem [{\citenamefont {Heifetz}\ \emph {et~al.}(2009)\citenamefont
  {Heifetz}, \citenamefont {Harnik},\ and\ \citenamefont
  {Tamarin}}]{heifetz2009canonical}%
  \BibitemOpen
  \bibfield  {author} {\bibinfo {author} {\bibfnamefont {E.}~\bibnamefont
  {Heifetz}}, \bibinfo {author} {\bibfnamefont {N.}~\bibnamefont {Harnik}},\
  and\ \bibinfo {author} {\bibfnamefont {T.}~\bibnamefont {Tamarin}},\
  }\bibfield  {title} {\bibinfo {title} {Canonical {H}amiltonian representation
  of pseudoenergy in shear flows using counter-propagating {R}ossby waves},\
  }\href@noop {} {\bibfield  {journal} {\bibinfo  {journal} {Q. J. Roy. Meteor.
  Soc.}\ }\textbf {\bibinfo {volume} {135}},\ \bibinfo {pages} {2161} (\bibinfo
  {year} {2009})}\BibitemShut {NoStop}%
\bibitem [{\citenamefont {Madelung}(1927)}]{madelung1927quantentheorie}%
  \BibitemOpen
  \bibfield  {author} {\bibinfo {author} {\bibfnamefont {E.}~\bibnamefont
  {Madelung}},\ }\bibfield  {title} {\bibinfo {title} {Quantentheorie in
  hydrodynamischer form},\ }\href@noop {} {\bibfield  {journal} {\bibinfo
  {journal} {Zeitschrift f{\"u}r Physik A Hadrons and Nuclei}\ }\textbf
  {\bibinfo {volume} {40}},\ \bibinfo {pages} {322} (\bibinfo {year}
  {1927})}\BibitemShut {NoStop}%
\bibitem [{\citenamefont {Heifetz}\ and\ \citenamefont
  {Cohen}(2015)}]{heifetz2015toward}%
  \BibitemOpen
  \bibfield  {author} {\bibinfo {author} {\bibfnamefont {E.}~\bibnamefont
  {Heifetz}}\ and\ \bibinfo {author} {\bibfnamefont {E.}~\bibnamefont
  {Cohen}},\ }\bibfield  {title} {\bibinfo {title} {Toward a
  thermo-hydrodynamic like description of schr{\"o}dinger equation via the
  madelung formulation and fisher information},\ }\href@noop {} {\bibfield
  {journal} {\bibinfo  {journal} {Foundations of Physics}\ }\textbf {\bibinfo
  {volume} {45}},\ \bibinfo {pages} {1514} (\bibinfo {year}
  {2015})}\BibitemShut {NoStop}%
\bibitem [{\citenamefont {Motsch}\ and\ \citenamefont
  {Tadmor}(2014)}]{motsch2014heterophilious}%
  \BibitemOpen
  \bibfield  {author} {\bibinfo {author} {\bibfnamefont {S.}~\bibnamefont
  {Motsch}}\ and\ \bibinfo {author} {\bibfnamefont {E.}~\bibnamefont
  {Tadmor}},\ }\bibfield  {title} {\bibinfo {title} {Heterophilious dynamics
  enhances consensus},\ }\href@noop {} {\bibfield  {journal} {\bibinfo
  {journal} {SIAM review}\ }\textbf {\bibinfo {volume} {56}},\ \bibinfo {pages}
  {577} (\bibinfo {year} {2014})}\BibitemShut {NoStop}%
\bibitem [{\citenamefont {Shvydkoy}\ and\ \citenamefont
  {Tadmor}(2018)}]{shvydkoy2018topological}%
  \BibitemOpen
  \bibfield  {author} {\bibinfo {author} {\bibfnamefont {R.}~\bibnamefont
  {Shvydkoy}}\ and\ \bibinfo {author} {\bibfnamefont {E.}~\bibnamefont
  {Tadmor}},\ }\bibfield  {title} {\bibinfo {title} {Topological models for
  emergent dynamics with short-range interactions},\ }\href@noop {} {\bibfield
  {journal} {\bibinfo  {journal} {arXiv preprint arXiv:1806.01371}\ } (\bibinfo
  {year} {2018})}\BibitemShut {NoStop}%
\bibitem [{\citenamefont {D{\"u}ring}\ \emph {et~al.}(2009)\citenamefont
  {D{\"u}ring}, \citenamefont {Markowich}, \citenamefont {Pietschmann},\ and\
  \citenamefont {Wolfram}}]{during2009boltzmann}%
  \BibitemOpen
  \bibfield  {author} {\bibinfo {author} {\bibfnamefont {B.}~\bibnamefont
  {D{\"u}ring}}, \bibinfo {author} {\bibfnamefont {P.}~\bibnamefont
  {Markowich}}, \bibinfo {author} {\bibfnamefont {J.-F.}\ \bibnamefont
  {Pietschmann}},\ and\ \bibinfo {author} {\bibfnamefont {M.-T.}\ \bibnamefont
  {Wolfram}},\ }\bibfield  {title} {\bibinfo {title} {Boltzmann and
  fokker--planck equations modelling opinion formation in the presence of
  strong leaders},\ }in\ \href@noop {} {\emph {\bibinfo {booktitle}
  {Proceedings of the Royal Society of London A: Mathematical, Physical and
  Engineering Sciences}}}\ (\bibinfo {organization} {The Royal Society},\
  \bibinfo {year} {2009})\ p.\ \bibinfo {pages} {rspa20090239}\BibitemShut
  {NoStop}%
\bibitem [{\citenamefont {Pedlosky}(1990)}]{pedlosky:1990}%
  \BibitemOpen
  \bibfield  {author} {\bibinfo {author} {\bibfnamefont {J.}~\bibnamefont
  {Pedlosky}},\ }\href@noop {} {\emph {\bibinfo {title} {{Geophysical Fluid
  Dynamics}}}},\ \bibinfo {edition} {2nd}\ ed.\ (\bibinfo  {publisher}
  {Springer-Verlag},\ \bibinfo {address} {New York, NY, USA},\ \bibinfo {year}
  {1990})\ p.\ \bibinfo {pages} {710}\BibitemShut {NoStop}%
\end{thebibliography}%

\end{document}